\documentclass[useAMS,usenatbib,fleqn]{mn2e}
\voffset=-0.7in \usepackage{amsmath}
\usepackage{dcolumn}
\usepackage{times}

\usepackage{graphicx}

                       \def\sun{\hbox{$\odot$}}

\def\lesssim{\mathrel{\hbox{\rlap{\hbox{\lower4pt\hbox{$\sim$}}}\hbox{$<$}}}}
\def\gtrsim{\mathrel{\hbox{\rlap{\hbox{\lower4pt\hbox{$\sim$}}}\hbox{$>$}}}}

\newcommand{\mamo}[1]{\mbox{$#1$}}
\newcommand{\unit}[1]{\ifmmode \:\mbox{\rm #1}\else \mbox{#1}\fi}

\newcommand{\sbr}[1]{_{\rm #1}}

\newcommand{\mone}{\mamo{^{-1}}}

\newcommand{\ten}[1]{\mamo{\times 10^{#1}}}

\newcommand{\kms}{\unit{km~s\mone}}
\newcommand{\kpc}{\unit{kpc}}
\newcommand{\mpc}{\unit{Mpc}}

\newcommand{\lcdm}{\mamo{\Lambda}CDM}
\newcommand{\secref}[1]{Section~\ref{sec:#1}}

\newcommand{\equref}[1]{equation~(\ref{eq:#1})}
\newcommand{\Eqref}[1]{Equation~(\ref{eq:#1})}
\newcommand{\figref}[1]{Fig.~\ref{fig:#1}}
\newcommand{\tabref}[1]{Table~\ref{tab:#1}}

\newcommand{\ugriz}{\emph{u*g'r'i'z'}}
\newcommand{\DelSig}{\mamo{\Delta \Sigma}}
\newcommand{\ur}{\mamo{u^* - r'}} 
\newcommand{\mstel}{\mamo{M_{*}}}
\newcommand{\fsat}{\mamo{f\sbr{sat}}}

\newcommand{\apj}{ApJ}
\newcommand{\apjl}{ApJL}
\newcommand{\apjs}{ApJS}
\newcommand{\mnras}{MNRAS}

\newcommand{\aap}{A\&A}
\newcommand{\aaps}{A\&AS}

\newcommand{\jcap}{JCAP}
\newcommand{\physrep}{Phys Rep}
\newcommand{\araa}{ARA\&A}

\newcommand{\lensfit}{{\em lens}fit}

\begin{document}

\title{CFHTLenS: Co-evolution of galaxies and their dark matter haloes}

\author[CFHTLenS]{\parbox{\textwidth}{
\vspace{-0.5cm}
\raggedright
\mbox{Michael~J.~Hudson\unskip$^{1,2}$\thanks{E-mail: mjhudson@uwaterloo.ca},}
\mbox{Bryan~R.~Gillis\unskip$^{1}$,}
\mbox{Jean~Coupon\unskip$^{3,4}$,}
\mbox{Hendrik~Hildebrandt\unskip$^{5}$,}
\mbox{Thomas~Erben\unskip$^{5}$,}
\mbox{Catherine~Heymans\unskip$^{6}$,}
\mbox{Henk~Hoekstra\unskip$^{7}$,}
\mbox{Thomas~D.~Kitching\unskip$^{8}$,}
\mbox{Yannick~Mellier\unskip$^{9,10}$,}
\mbox{Lance~Miller\unskip$^{11}$,}
\mbox{Ludovic~Van~Waerbeke\unskip$^{12}$,}
\mbox{Christopher Bonnett\unskip$^{13}$,}
\mbox{Liping~Fu\unskip$^{14}$,}
\mbox{Konrad~Kuijken\unskip$^{7}$,}
\mbox{Barnaby~Rowe\unskip$^{15}$,}
\mbox{Tim~Schrabback\unskip$^{16, 5}$,}
\mbox{Elisabetta~Semboloni\unskip$^{7}$,}
\mbox{Edo~van~Uitert\unskip$^{7,5}$,}
\mbox{Malin~Velander\unskip$^{11,7}$}
}
\vspace{0.4cm}\\
\parbox{\textwidth}{
$^1$ Department of Physics \& Astronomy, University of Waterloo,
Waterloo, ON, N2L 3G1, Canada.\\
$^2$ Perimeter Institute for Theoretical Physics, 31 Caroline St. N.,
Waterloo, ON, N2L 2Y5, Canada.\\
$^{3}$ Institute of Astronomy and Astrophysics, Academia Sinica, P.O. Box 23-141, Taipei 10617, Taiwan.\\
$^{4}$ Department of Astronomy, University of Geneva, ch. d'Ecogia 16,
CH-1290 Versoix, Switzerland.\\
$^{5}$ Argelander Institute for Astronomy, University of Bonn, Auf dem H{\"u}gel 71, 53121 Bonn, Germany.\\
$^{6}$ Scottish Universities Physics Alliance, Institute for Astronomy, University of Edinburgh, Royal Observatory, Blackford Hill, Edinburgh, EH9 3HJ, U.K.\\
$^{7}$ Leiden Observatory, Leiden University, Niels Bohrweg 2, 2333 CA Leiden, The Netherlands.\\
$^{8}$ Mullard Space Science Laboratory, University College London, Holmbury St Mary, Dorking, Surrey RH5 6NT, U.K.\\
$^{9}$ Institut d'Astrophysique de Paris, Universit\'{e} Pierre et Marie Curie - Paris 6, 98 bis Boulevard Arago, F-75014 Paris, France.\\
$^{10}$ Institut d'Astrophysique de Paris, CNRS, UMR 7095, 98 bis Boulevard Arago, F-75014 Paris, France.\\
$^{11}$ Department of Physics, Oxford University, Keble Road, Oxford OX1 3RH, U.K.\\
$^{12}$ University of British Columbia, Department of Physics and Astronomy, 6224 Agricultural Road, Vancouver, B.C. V6T 1Z1, Canada.\\
$^{13}$ Institut de Ciencies de l'Espai, CSIC/IEEC, F. de Ciencies, Torre C5 par-2, Barcelona 08193, Spain.\\
$^{14}$ Shanghai Key Lab for Astrophysics, Shanghai Normal
University, 100 Guilin Road, 200234, Shanghai, China.\\
$^{15}$ Department of Physics and Astronomy, University College London, Gower Street, London WC1E 6BT, U.K.\\
$^{16}$ Kavli Institute for Particle Astrophysics and Cosmology, Stanford University, 382 Via Pueblo Mall, Stanford, CA 94305-4060, U.S.A.\\
\vspace{-0.5cm}}
}

\maketitle

\begin{abstract}
Galaxy-galaxy weak lensing is a direct probe of the mean matter distribution around galaxies. The depth and sky coverage of the CFHT Legacy Survey yield statistically significant galaxy halo mass measurements over a much wider range of stellar masses ($10^{8.75}$ to $10^{11.3} M_{\sun}$) and redshifts ($0.2 < z < 0.8$) than previous weak lensing studies. At redshift $z \sim 0.5$, the stellar-to-halo mass ratio (SHMR) reaches a maximum of $4.0\pm0.2$ percent as a function of halo mass at $\sim 10^{12.25} M_{\sun}$. We find, for the first time from weak lensing alone, evidence for significant evolution in the SHMR: the peak ratio falls as a function of cosmic time from $4.5 \pm 0.3$ percent at $z \sim 0.7$ to $3.4 \pm 0.2$ percent at $z \sim 0.3$, and shifts to lower stellar mass haloes. These evolutionary trends are dominated by red galaxies, and are consistent with a model in which the stellar mass above which star formation is quenched ``downsizes'' with cosmic time. In contrast, the SHMR of blue, star-forming galaxies is well-fit by a power law that does not evolve with time. This suggests that blue galaxies form stars at a rate that is balanced with their dark matter accretion in such a way that they evolve along the SHMR locus. The redshift dependence of the SHMR can be used to constrain the evolution of the galaxy population over cosmic time.
\end{abstract}

\begin{keywords}
cosmology: observations -- gravitational lensing: weak -- galaxies: haloes -- dark matter \vspace{-0.55cm}
\end{keywords}

\section{Introduction}
\label{sec:intro}

A full understanding of the co-evolution of the stellar, gaseous and dark matter (DM) components of galaxies, and the physical causes thereof, is the primary goal of studies of galaxy formation and evolution.  Observations directly yield a snapshot of the luminosities (and stellar masses) of the galaxy population at a given redshift, whereas numerical studies  most easily predict the abundance and evolution of DM haloes. One way to connect these two is with the ``abundance matching'' ansatz: galaxies are assigned to DM haloes by rank-ordering each set and matching them one-to-one from highest to lowest. Using this assumption, \cite{MarHud02} showed that galaxy formation is most efficient in haloes of mass $\sim 10^{12.5} M_{\sun}$, at which mass 25\% of the baryons had been converted to stars. The method of \cite{MarHud02} was based on the summed luminosity of all galaxies in a halo, and was then extended to consider central galaxies and satellites \citep{YanMovan03, ValOst04, ConWecKra06, MosSomMau10}.   Another approach is to populate haloes with galaxies (the so-called ``Halo Occupation Distribution'', or HOD) in order to reproduce galaxy clustering \citep{JinMoBoe98, Sel00, PeaSmi00, BerWei02, CooShe02, KraBerWec04, ZehZheWei05, CouKilMcC12}.  While these methods are powerful, they are model-dependent in the sense that some critical assumptions are made in the statistical link between galaxies and their haloes. 

Connecting the galaxies to their dark matter haloes in a more direct fashion requires probes of the gravitational effects of the DM haloes. There are several ways to measure galaxy masses at intermediate redshifts. Traditionally, dynamical methods have been used to obtain masses. All dynamical methods make some assumption regarding the dynamical equilibrium of the system. Furthermore, some methods, such as the Tully-Fisher \citep{TulFis77} relation or the Fundamental Plane \citep{DjoDav87} only probe the inner regions of the halo. Other methods, such as satellite kinematics \citep[e.g.][]{MorvanCac11}, reach further out in the halo but are difficult to apply at intermediate redshifts due to the faintness of the satellites.

A powerful alternative approach to these dynamical methods is to use weak gravitational lensing to measure the masses of galaxy DM haloes \citep{BraBlaSma96}. Weak lensing is sensitive to all of the matter that surrounds the galaxy and along the line of sight. Because galaxy-galaxy lensing (GGL) is an ensemble mean measurement, this includes matter that is statistically correlated with galaxy haloes: in other words, GGL is measuring the galaxy-matter cross-correlation function. This has led to the development of a ``halo model'' \citep[e.g.][hereafter M06]{ManSelKau06} to interpret the GGL measurements in a similar way as had been done in studies of galaxy clustering. The halo model has been applied to recent weak lensing data sets from the RCS2 by \cite{vanHoeVel11}, and the CFHT Lensing Survey (CFHTLenS) by \citet[hereafter V14]{VelvanHoe14}.

The focus of this paper is to use GGL to study galaxy evolution, or more specifically, the evolution with redshift of the stellar-to-halo-mass ratio (hereafter SHMR).  In this sense, this paper parallels recent efforts to extend the abundance matching and HOD methods to higher redshifts. The promise of using GGL to probe galaxy evolution goes back to \cite{HudGwyDah98}, who found no relative evolution between the GGL signal in the Hubble Deep Field and the low-redshift Tully-Fisher and Faber-Jackson relations.  \citet[hereafter L12]{LeaTinBun12} performed a combined HOD analysis using GGL, abundance matching and correlation functions on COSMOS data. They found that the peak value of the SHMR did not evolve with redshift, but that SHMR ``downsizes'' in the sense that the halo (and stellar) mass at which it peaks decreases with cosmic time. The RCS2 GGL study of \cite{vanHoeVel11} examined the evolution of the SHMR but lacked the statistical power to place strong constraints. V14 analyzed GGL in the CFHTLenS, but limited to the redshift range $0.2 < z < 0.4$. The combination of depth and area of the CFHTLenS sample allows us, for the first time, to split lens samples by redshift, colour and stellar mass, and hence to measure the evolution of the SHMR using only GGL.

The outline of this paper is as follows. We discuss the CFHTLenS shape and photometric redshift (hereafter photo-$z$) data in \secref{data}, and describe the halo model in \secref{models}. In \secref{results}, we show the fits of the halo model to the weak lensing data. The SHMR is discussed in \secref{shmr}, and the GGL results are compared with SHMR results from other methods. In \secref{discuss}, we discuss the interpretation of the SHMR in terms of models of star formation and quenching. We also compare our results for faint blue galaxies to determinations from galaxy rotation curves. Our conclusions are summarised in \secref{conc}.

Throughout we adopt a `737' \lcdm\ cosmology: a Hubble parameter, $h \equiv H_0/(100 \kms \mpc^{-1}) = 0.7$, a matter density parameter $\Omega\sbr{m,0} = 0.3$ and a cosmological constant $\Omega_{\Lambda,0} = 0.7$ The values of $\Omega\sbr{m,0}$ and   $\Omega_{\Lambda,0}$ are consistent with the current best-fit WMAP9 cosmology. \citep[including Baryonic Acoustic Oscillations and $H_{0}$]{HinLarKom13} as well as with the first Planck results \citep[including BAO]{Planck13-16}. There is a well known tension between the Planck value of $h$ and that derived by \cite{RieMacCas11}, the value adopted here lies in between these. All masses, distances and other derived quantities are calculated using this value of $h$.

\section{Data}
\label{sec:data}

The data used in this paper is based on the ``Wide Synoptic'' and ``Pre-Survey'' components of the Canada-France-Hawaii Telescope Legacy Survey (CFHTLS),  a joint project between Canada and France.  The CFHTLenS collaboration analysed these data and produced catalogues of photometry, photometric redshifts and galaxy shapes as described below.

\subsection{Images and photometry}

The survey area was imaged using the Megaprime wide field imager mounted at the prime focus of the Canada-France-Hawaii Telescope (CFHT). The MegaCam camera is an array of $9\times4$ CCDs with a field of view of 1~deg$^2$. The CFHTLS wide synoptic survey covers an effective area of 154~deg$^2$ in five bands: $u^*$, $g'$, $r'$, $i'$ and $z'$. This area consists of four independent fields, W1--4 
with a full multi-colour depth of $i'_{\mathrm{AB}}=24.7$ (for a source in the CFHTLenS catalogue). The images have been independently reduced within the CFHTLenS collaboration; the details of the data reduction are described in detail in \citet{ErbHilMil13}. 

\subsection{Source shapes}

CFHTLenS has measured shapes for $8.7\times10^6$ galaxies \citep[][]{HeyvanMil12,MilHeyKit13} with $i'_{AB} < 24.7$  with the \lensfit\ algorithm \citep{MilKitHey07, KitMilHey08, MilHeyKit13}. These have been thoroughly tested for systematics within the CFHTLenS collaboration \citep[see][]{HeyvanMil12}. The ellipticities have a  Gaussian scatter of $\sigma_{e} = 0.28$ \citep{HeyGroHea13}. The ellipticities are almost unbiased estimates of the gravitational shear: there is a small multiplicative correction discussed in \secref{stacking} below. We do not apply the weak additive $c$-term correction discussed in \citet{HeyvanMil12} as we found that it had no effect on our GGL results (V14).

\subsection{Photometric redshifts and stellar masses}

The CFHTLenS photometric redshifts, $z\sbr{p}$, are described in detail in \citet{HilErbKui12}. Over the redshift range of interest for this paper, these photo-$z$'s are typically precise to $\pm0.04(1+z)$, with a 2-5\% catastrophic failure rate. The photo-$z$ code also fits a spectral template, ranging from $T = 1$ (elliptical) to $T = 6$ (starburst).  Here we correct the photometric redshifts for small biases with respect to spectroscopic redshifts, as discussed in more detail in Appendix \ref{sec:photozbias}.

Stellar masses, \mstel, are measured using the LePhare \citep{IlbArnMcC06} code, with the photometric redshift fixed at the value found by \citet{HilErbKui12}. The \ugriz\ magnitudes are fit using \cite{BruCha03} models with varying exponential star formation histories and dust extinction, as described in more detail in V14. That paper also compared the stellar masses from CFHTLenS \ugriz\ photometry to those determined using the same code but with \ugriz\ plus infrared photometry (based on WIRDS data).  There are slight systematic differences between WIRDS and CFHTLenS stellar masses (for both red and blue galaxies) amounting to $0.1-0.2$ dex.  
The LePhare fits also produce absolute magnitudes in all Megaprime bands. We will use the rest-frame $u^* - r'$ colours to separate red and blue galaxies in \secref{lenses} below.

\subsection{Sample selection}

\begin{figure*}
\begin{center}
\includegraphics[width=\textwidth]{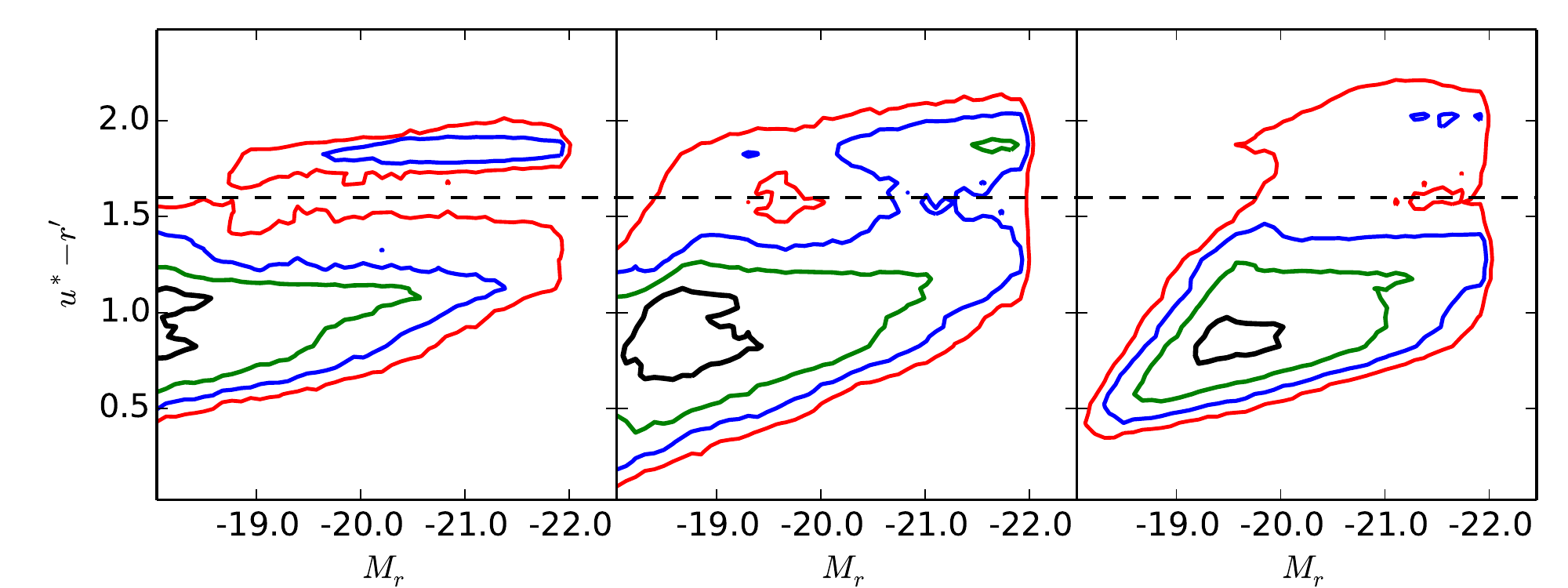}
\caption{Colour-magnitude diagrams for CFHTLenS galaxies for three redshift bins $0.2 \le z\sbr{p} < 0.4$, $0.4 \le z\sbr{p} < 0.6$, $0.6 \le z\sbr{p} < 0.8$ from left to right. The contours show density of galaxies in colour-magnitude space relative to the peak density in that redshift bin. The colour is restframe $u*\sbr{CFHT}-r'\sbr{CFHT}$, the horizontal dashed line shows the colour 1.6 used to separate red and blue populations.}
\label{fig:urz}
\end{center}
\end{figure*}

\subsubsection{Lenses}
\label{sec:lenses}

We select lenses in the range $0.2 < z\sbr{p} < 0.8$ with $i\sbr{AB}' < 23$. We use the Megaprime rest-frame \ur\ colours to separate red and blue galaxies, with the division at $\ur = 1.60$, independent of magnitude or redshift, as we observe no strong evolution in the red/blue division. 
The colour magnitude diagram for lenses in three different redshift bins is shown in \figref{urz}.
This criterion is similar to, but not identical to that of V14, who used the fitted spectral template type $T$ to separate the red and blue populations. In this paper we choose to use colour 
to make subsequent comparisons with our results more straightforward.
This corresponds approximately to $u\sbr{SDSS}'-r\sbr{SDSS}' = 1.87$, similar to the colour  $u\sbr{SDSS}'-r\sbr{SDSS}' = 1.82$ used by \citep{BalGlaBri04} to separate red and blue galaxies at faint luminosities in the Sloan Digital Sky Survey (SDSS).

In addition to the red/blue colour separation discussed above, we will bin lenses by redshift ($0.2 \le z\sbr{p} < 0.4$, $0.4 \le z\sbr{p} < 0.6$, $0.6 \le z\sbr{p} < 0.8$). With these cuts there are 1.62\ten{6} blue galaxies and 4.50\ten{5} red galaxies, for a total of 2.06\ten{6} lenses. We will also bin lenses by $r'$-band luminosity/stellar mass. However, the stellar masses of individual galaxies are noisy. Were we to bin by stellar mass, the noise would introduce Eddington-like biases, and this would require simulations to correct (as in e.g.\ V14). Instead, we bin by $r$-band luminosity and use the mean stellar-mass-to-light ratio for that bin to calculate the mean stellar mass of the bin. 
The $r'$-band magnitude limits for the bins are chosen for red and blue galaxies separately so that the stellar mass bins are approximately 0.5 dex in width, for $\mstel > 10^{8.5} M_{\sun}$.

\subsubsection{Sources}
\label{sec:sources}

Sources are limited to $i\sbr{AB}' < 24.7$ from unmasked regions of the CFHTLS.  We use the full unmasked survey area, including the fields which did not pass cosmic shear systematics tests described in \cite{HeyvanMil12}. V14 found that, for GGL, there was no difference between results from these fields and the remainder of the survey. We also limit the source redshifts to $z\sbr{p} < 1.3$, where the photo-$z$s are reliable \citep{HilErbKui12}.  Excluding masked areas, there are $5.6\times10^6$ sources, corresponding to an effective source density of $10.6\,\mathrm{arcmin}^{-2}$ \citep{HeyvanMil12}.  

\subsubsection{Lens-source pairs}
\label{sec:pairs}

From the lens and source  samples, we analyse all lens-source pairs with $\Delta z\sbr{p} > 0.1$. At the median redshift of the lenses $z\sbr{l} \sim 0.5$, the typical error in source redshift is $\sim 0.05$, hence this yields a $\sim 2 \sigma$ separation in redshift space. Note that later we will downweight close lens-source pairs, with the result that any physical pairs in which one member is scattered up by photo-$z$ errors will have low weight in any case.

We also only consider lens-source pairs that are not too close on the sky. There are two reasons for this. First, the shapes of background sources may be affected by the extended surface brightness profile of bright foreground galaxies \citep{HudGwyDah98, VelKuiSch11}. Schrabback et al.\ (in prep.) have examined this issue via simulated galaxy catalogues including realistic surface brightness distributions. They find no additive bias for pairs as close as 3 arcseconds.  This angular separation corresponds to 10 kpc at $z = 0.2$ or 22.5 kpc at $z=0.8$. Second, it is possible that features within a galaxy (e.g.\ spiral arms) may be split by the SExtractor \citep{BerArn96} software and treated as independent ``sources.''   These fragments will have assigned photo-$z$s, and our cut $\Delta z\sbr{p} > 0.1$ should remove most of these spurious pairs.  There remains, however, the possibility of residual contamination, and so a cut based on the luminosity profiles of bright disks is also applied. \citet{vanFre11} find that spiral galaxy disks typically truncate at an isophote corresponding to a $B$-band surface brightness of 26 mag.\ per square arc second. The largest discs have low surface brightness (SB); \cite{AllDriGra06} find few disks with surface brightness (measured at the effective radius) fainter than 24 in $B$.  To be conservative, we use an SB limit of 27 for the isophotal truncation and faint limit of 25 in ${B}$ for the lowest surface brightness galaxies. This yields an estimated truncation radius of $R\sbr{trunc} = 37 \kpc$ for a $B$-band fiducial magnitude of $-20.5$ (or an $r'$-band magnitude of $-21.36$), and a scaling $R\sbr{trunc} \propto (L/L\sbr{fid})^{0.5}$.
We adopt double this for the minimum radius, $R\sbr{min}$,  for lens-source pairs around blue galaxies, but with a hard lower limit of 20 \kpc\ and a hard upper limit of 50 \kpc. For red galaxies, we adopt the same minimum radius as for blue galaxies.  In practice, then this is always larger than the 3 arcsecond angular cut discussed above. For each bin of redshift and luminosity we adopt the larger of the two radii as the minimum projected radius of the lens-source pairs.

\subsection{Stacking and weights}
\label{sec:stacking}

The excess surface density,
\begin{equation}
\Delta\Sigma(R) = \overline{\Sigma(<R)}-\overline{\Sigma(R)}\textrm{,}
\label{eq:mainlens}
\end{equation}
defined as the difference between $\overline{\Sigma(<R})$, the mean projected surface mass density enclosed within a circle of radius $R$ and $\overline{\Sigma(R)}$, the average surface density at radius $R$, can be related to the observed tangential shear $\gamma_{t}$ via
\citep{Mir91, FahKaiSqu94}:
\begin{equation}
\Delta\Sigma(R) = \Sigma_{\rm crit}\langle\gamma_t(R)\rangle \,,
\label{eq:surfaceDensity}
\end{equation}
where critical surface density, $\Sigma_{\rm crit}$, is given by
\begin{equation}
\Sigma_{\rm crit} = \frac{c^2}{4\pi G}\frac{D\sbr{s}}{D\sbr{l} D\sbr{ls}} \,,
\label{cfhtls:eq:criticalSurfaceDensity}
\end{equation}
and where $D\sbr{l}$ is the angular diameter distance to the lens, $D\sbr{s}$ is the angular diameter distance to the source and $D\sbr{ls}$ is the angular diameter distance between lens and source. 

It is necessary to stack a large number of lenses to obtain a statistically significant signal. We calculate the surface mass density as a function of projected separation, $R$, by summing the tangential component of the source ellipticities over all lens-source pairs. We weight the sources by their \lensfit\ weights, $w$, which includes both the ellipticity measurement error and the intrinsic shape noise \citep[eq. 8]{MilHeyKit13}. We also weight pairs by $W = \Sigma\sbr{crit}^{{-2}}$. The excess surface density is then given by
\begin{equation}
\langle\DelSig(R)\rangle = \frac{\sum w_j e_{t,j} \Sigma_{\textrm{crit},ij}W_{ij}}{\sum w_j W_{ij}}
\end{equation}
where the sum is over all pairs of lenses, $i$, and sources, $j$, in a given $R$ bin, and $e_{t,j}$ is the tangential ellipticity of the source.
As in V14, we correct the ellipticities for a small bias in the \lensfit\ method: a calibration factor $m(\nu_{\mathrm{SN}},r_{\mathrm{gal}})$, which is modelled as a function of the signal-to-noise ratio, $\nu_{\mathrm{SN}}$, and size of the source galaxy, $r_{\mathrm{gal}}$ as described in \cite{MilHeyKit13}. Rather than dividing each galaxy shear by a factor $(1+m)$, which would lead to a biased calibration as discussed in \citet{MilHeyKit13}, we apply it to our average shear measurement as a function of lens redshift using the correction
\begin{equation}
1+K(z\sbr{lens}) = \frac{\sum w_j W_{j}[1+m(\nu_{\mathrm{SN},j},r_{\mathrm{gal},j})]}{\sum w_j W_{j}}\;.
\end{equation}
The lensing signal is then calibrated as follows:
\begin{equation}
\langle\DelSig^{\mathrm{cal}}\rangle = \frac{\langle\DelSig\rangle}{1+K(z\sbr{lens})}\;.
\label{eq:corrfac}
\end{equation}
The effect of this correction term on our GGL analysis is to increase the average lensing signal amplitude by 6.5\% at $z\sbr{lens} = 0.2$. As the lens redshift increases, the sources behind it become fainter and smaller and so the correction rises to 9\% at $z\sbr{lens} = 0.8$. 

The scatter in the photo-$z$s of lenses and sources will also introduce a bias in quantities such as the redshift and luminosity of the lens and the distance ratio $D\sbr{ls}/D\sbr{s}$. Appendix \ref{sec:photozscatter} discusses how we use simulations of mock catalogues to estimate these biases. These bias estimates are then used to correct all affected quantities.

\section{Models}
\label{sec:models}

\subsection{Halo model}
\label{sec:halo}

We will fit the data with simple halo models that describe the average distribution of total matter around a given galaxy, i.e.\ the matter-galaxy cross-correlation $\DelSig(R)$.  This can be broken into several terms: the first arises from the galaxy's own stellar mass and halo, or subhalo if it is a satellite. The second term is the ``offset group-halo'' term 
and is the mean distribution of DM around a given satellite galaxy due to the host halo that the satellite inhabits\footnote{The offset-group term is referred to as the ``satellite one-halo'' term by M06 and others, but we find this can be confused with the subhalo's one-halo term.}. The third term, which we neglect here because it is important only on very large ($\gtrsim 1000$ kpc) scales, is the two-halo term, which represents the matter in separate haloes that are correlated with the host halo.  This yields:
\begin{equation}
\DelSig(R) = \DelSig\sbr{1h}(R) + \DelSig\sbr{OG}(R)\end{equation}
We now describe each term in more detail.
\subsubsection{One-halo term}
\label{sec:onehalo}

The so-called one-halo term arises from matter within the galaxy's own halo:  its stars, \mstel, and its DM halo.  The stellar mass is modelled as a point source:
\begin{equation}
\DelSig_{*}(R) = \mstel/(\pi R^2)
\end{equation}
The DM is modelled as a \citet[hereafter NFW]{NavFreWhi97} density profile, parametrized by a virial mass, $M_{200}$, defined within the radius $r_{200}$ enclosing a mean density 200 times the critical density and a concentration $c_{200} = r_{200}/r\sbr{s}$ with $r\sbr{s}$ being the NFW scale radius. The concentration of the DM halo, $c_{200}$, is not free, but instead is fixed as a function of $M_{200}$ and redshift $z$ using the relation given by \cite{MunMacGot11}, converted from $(M\sbr{vir}, c\sbr{vir})$ to $(M_{200}, c_{200})$ using the method of \cite{HuKra03}. The excess surface density  $\DelSig\sbr{NFW}$ for an NFW profile is given in \cite{BalMarOgu09}.

While the NFW profile has been shown to be a good description of isolated haloes, the DM haloes of satellite galaxies are expected to be tidally stripped by their ``host'' DM halo \citep{TayBab01}. This effect has been observed in clusters of galaxies by weak lensing \citep{LimKneBar07, NatKneSma02}. Weak lensing has also been used to detect this effect statistically within groups and clusters in the CFHTLenS sample itself by \cite{GilHudErb13}. They found that, on average, satellites in high density environments had $35\pm12\%$ of their mass stripped, or, equivalently, were stripped to a truncation radius of  $(0.26\pm0.14)\times r_{200}$. Here we adopt a truncation radius $R_t = 0.4\, r_{200}$, which is consistent with the \cite{GilHudErb13} result, but which allows a straightforward comparison with the results of  previous authors (M06,V14).  For this truncated NFW profile (tNFW), we assume $\DelSig(R) \propto R^{-2}$ beyond the truncation radius:
\begin{equation}
\DelSig\sbr{tNFW}(R) = 
\begin{cases}
\DelSig\sbr{NFW}(R), & R \le R_t \\
\DelSig\sbr{NFW}(R_t)\times  \left(\frac{R_t^2}{R^2}\right) & R > R_t
\end{cases}
\label{eq:trunc}
\end{equation}
The effect of this assumption is examined in greater detail in Appendix \ref{sec:subhalo}.
 
In each galaxy subsample as binned by mass, colour and redshift, a fraction, \fsat, of the galaxies will be satellites with the remainder being ``central" galaxies.  Thus the one-halo term consists of the stellar mass, plus a mean (weighted by satellite fraction) of the central and stripped satellite dark haloes. Note that because of the stripping prescription, for satellites, the fitted value of the parameter $M_{200}$ is the mass before they fell into their host halo. In summary,
\begin{multline}
\DelSig\sbr{1h} = \DelSig_{*} + 
(1-\fsat)\DelSig\sbr{NFW}(M_{200}, c) \\
+ \fsat \DelSig\sbr{tNFW}(M_{200}, c) \,.
\end{multline}
The total one-halo mass, $M\sbr{h}$ is the sum of the baryons and the NFW $M_{200}$. 

\subsubsection{Offset group halo}
\label{sec:offsetgroup}

On intermediate scales (200 --- 1000 kpc),  the ``offset-group-halo" term dominates the \DelSig\ signal. This term is given by eqs. 11-13 of \cite{GilHudErb13}. In brief, this arises for satellite galaxies only and is due to the DM in their host halo.  It is a convolution of the projected NFW profile with the distribution of satellites, and so it depends on the mass of the host group, the concentration, $c$, of the DM in that group halo and, because it is a convolution of satellite positions, it also depends on the radial distribution of the satellites with respect to group centre. As with the 1-halo term, we assume that the hosting DM haloes have concentrations given by the prescription of \cite{MunMacGot11}. The distribution of satellite host halo masses is taken from the halo model of \cite{CouKilMcC12}. Finally, we assume that the concentration of satellites, $c\sbr{sat}$, is the same as that of the DM, which is consistent with the assumptions made by V14 and \cite{CouKilMcC12}.  
The satellite fraction, $f\sbr{sat}$ is constrained by the offset-group term.

\subsection{Fitting the halo model} 
\label{sec:fitting}

In addition to the predicted $\DelSig(R)$, a full treatment of the halo model also contains a detailed prescription of how galaxies occupy haloes as a function of their magnitude and as a function of the halo mass.  V14 adopt a halo model in which the parameters of the offset-group term are coupled to the one halo term. The approach taken here is somewhat different: for satellites, we will adopt the HOD parameters from \cite{CouKilMcC12}. This then  specifies the distribution of host halo masses for a given satellite stellar mass and hence the \emph{shape} of the offset-group term. The model for the one-halo term is thus independent of that of the offset-group term. This leaves only two free parameters: $M_{200}$ of the one-halo term; and the satellite fraction, $f_{sat}$.

In practice, there is some degeneracy between the satellite fraction and the one-halo mass.  \cite{CouKilMcC12} show that there is little evolution in the satellite fraction in their HOD fits and that it is consistent with a linear function in magnitude (or, equivalently, log stellar mass). We adopt these constraints, and fit a non-evolving linear satellite fraction.

\section{Results}
\label{sec:results}

\begin{figure*}
\begin{center}
\includegraphics[width=\textwidth]{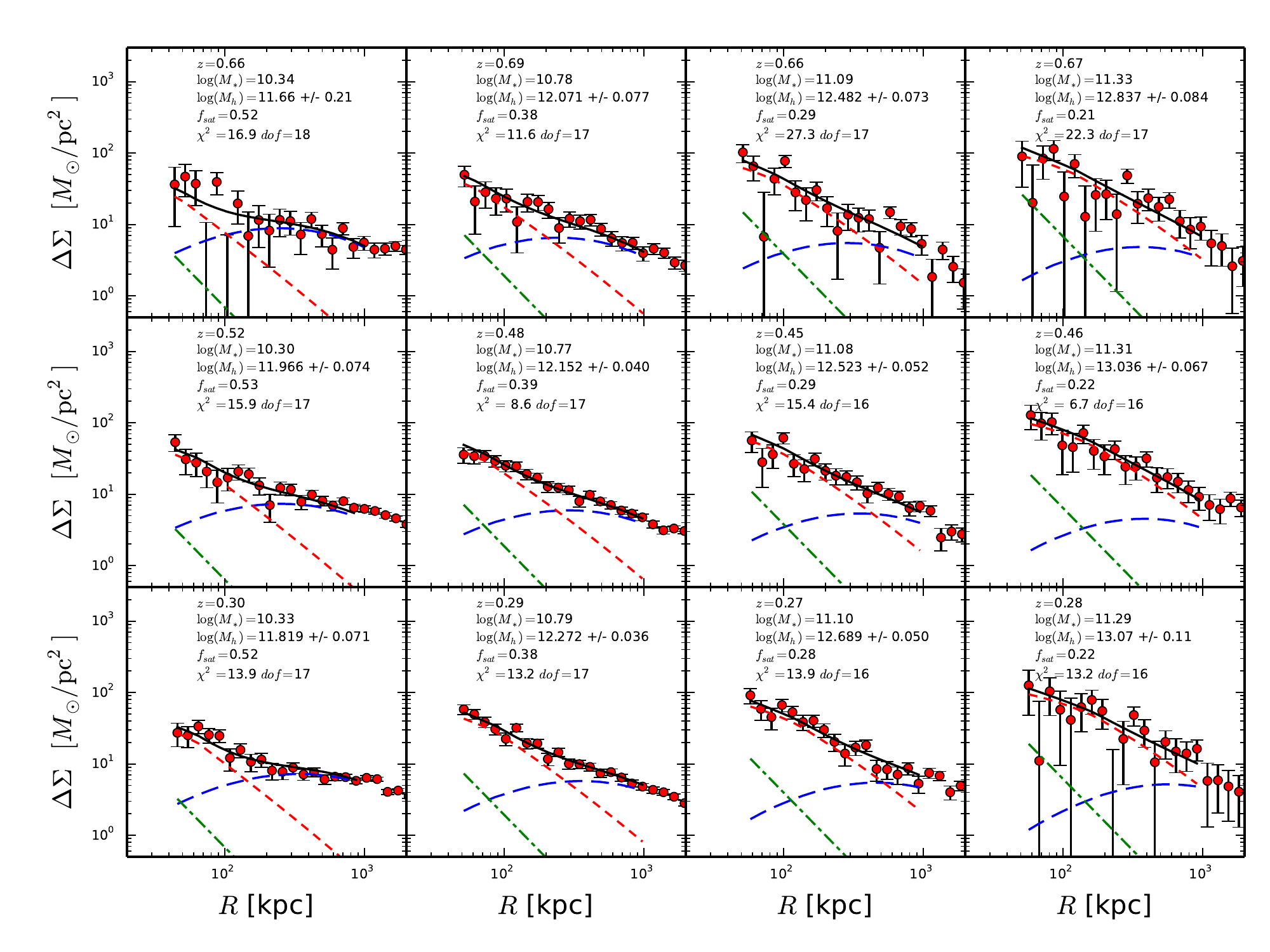}
\caption{\DelSig\ as a function of projected radius for red galaxies. Each panel shows a specific data from a bin in stellar mass and redshift, with redshift decreasing from top ($z \sim 0.7$) to bottom ($z \sim 0.3$) and stellar mass increasing from left to right. The points show the CFHTLenS data. Model fits show the NFW halo (red short-dashed), stellar mass (green dash-dot) and offset group halo term (blue long-dashed). The sum is plotted in black. Data are plotted to 2000 kpc, but fits are performed using only data to 1000 \kpc.}
\label{fig:delsigred}
\end{center}
\end{figure*}

\begin{figure*}
\begin{center}
\includegraphics[width=\textwidth]{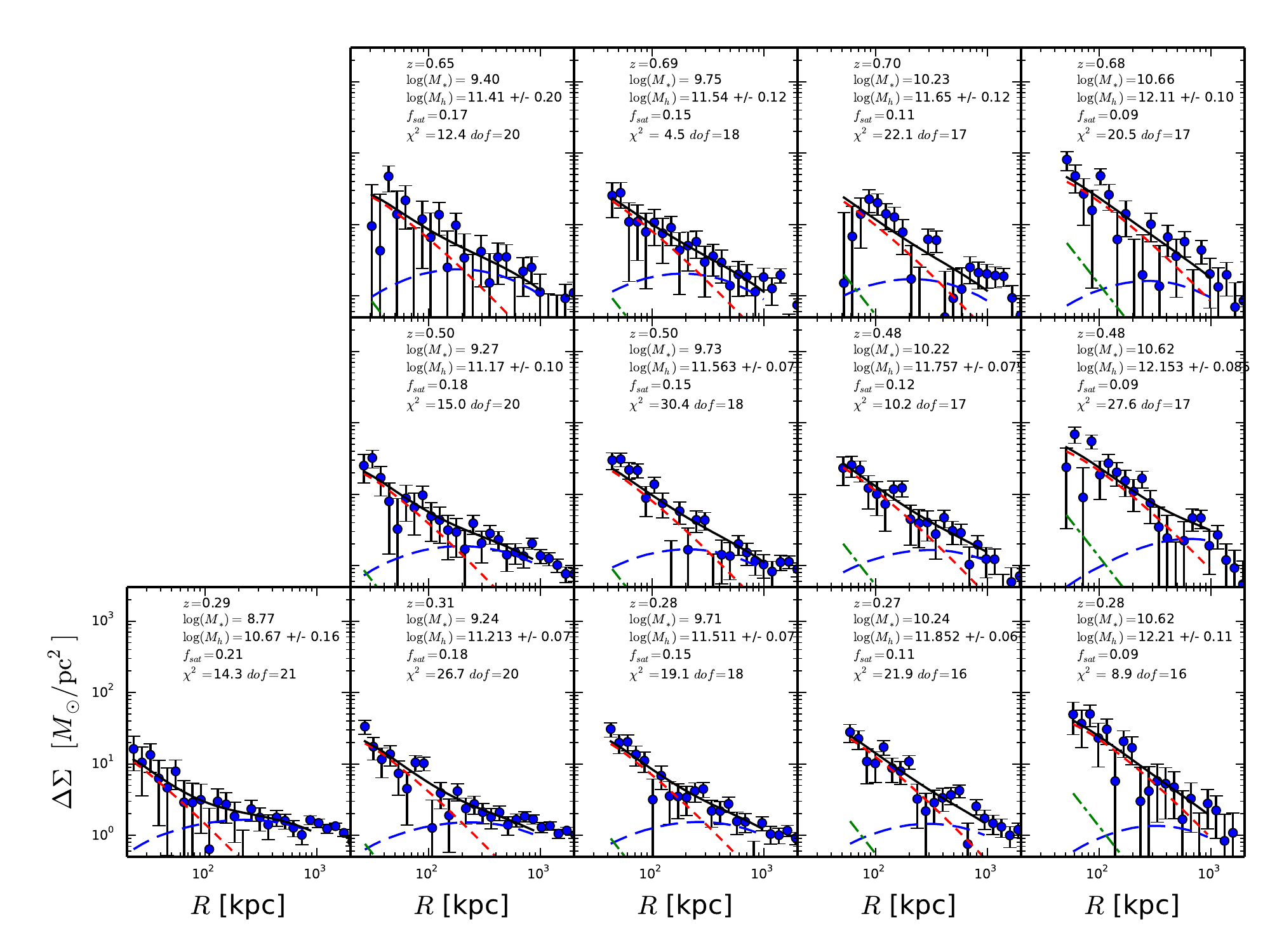}
\caption{As in \figref{delsigred} but for blue galaxies}
\label{fig:delsigblue}
\end{center}
\end{figure*}

\makeatletter{}\begin{table*}
\newcolumntype{d}[1]{D{.}{.}{#1} }
\caption{Halo model fits for red galaxies}
\begin{tabular}{rd{-1}d{-1}d{-1}d{-1}ccd{-1}r}
\multicolumn{1}{c}{$N\sbr{lens}$} & \multicolumn{1}{c}{$\langle z\sbr{lens} \rangle$} & \multicolumn{1}{c}{$\langle u_*-r \rangle$} & \multicolumn{1}{c}{$\langle M_r \rangle$} & \multicolumn{1}{c}{$M_*$} & \multicolumn{1}{c}{$M\sbr{h}$} & \multicolumn{1}{c}{$f\sbr{sat}$} & \multicolumn{1}{c}{$\chi^2$} & \multicolumn{1}{c}{d.o.f.} \\
 &  &  &  & \multicolumn{1}{c}{$10^{10} M_{\sun}$} & \multicolumn{1}{c}{$10^{11} M_{\sun}$} &  &  &  \\
41280 & 0.30 & 1.81 & -20.67 & 2.14 & $6.6 \pm 1.1$ & $0.523 \pm 0.016$ & 13.9 & 17 \\
33620 & 0.28 & 1.84 & -21.71 & 6.17 & $18.7 \pm 1.6$ & $0.380 \pm 0.012$ & 13.2 & 17 \\
4115 & 0.27 & 1.86 & -22.60 & 12.59 & $48.8 \pm 5.6$ & $0.283 \pm 0.009$ & 13.9 & 16 \\
304 & 0.28 & 1.87 & -23.10 & 19.50 & $117.0 \pm 29.0$ & $0.223 \pm 0.007$ & 13.2 & 16 \\
66170 & 0.52 & 1.84 & -20.60 & 2.00 & $9.3 \pm 1.6$ & $0.532 \pm 0.017$ & 15.9 & 17 \\
94590 & 0.48 & 1.87 & -21.67 & 5.89 & $14.2 \pm 1.3$ & $0.385 \pm 0.012$ & 8.6 & 17 \\
13840 & 0.45 & 1.89 & -22.55 & 12.02 & $33.3 \pm 4.0$ & $0.289 \pm 0.009$ & 15.4 & 16 \\
1941 & 0.46 & 1.91 & -23.15 & 20.42 & $109.0 \pm 17.0$ & $0.217 \pm 0.007$ & 6.7 & 16 \\
48690 & 0.66 & 1.90 & -20.69 & 2.19 & $4.5 \pm 2.2$ & $0.520 \pm 0.016$ & 16.9 & 18 \\
109100 & 0.69 & 1.95 & -21.69 & 6.03 & $11.8 \pm 2.1$ & $0.382 \pm 0.012$ & 11.6 & 17 \\
28820 & 0.66 & 1.95 & -22.57 & 12.30 & $30.3 \pm 5.1$ & $0.286 \pm 0.009$ & 27.3 & 17 \\
7161 & 0.67 & 1.98 & -23.21 & 21.38 & $69.0 \pm 13.0$ & $0.210 \pm 0.006$ & 22.3 & 17 \\
\end{tabular}
\label{tab:redfits}
\end{table*}

\makeatletter{}\begin{table*}
\newcolumntype{d}[1]{D{.}{.}{#1} }
\caption{Halo model fits for blue galaxies}
\begin{tabular}{rd{-1}d{-1}d{-1}d{-1}ccd{-1}r}
\multicolumn{1}{c}{$N\sbr{lens}$} & \multicolumn{1}{c}{$\langle z\sbr{lens} \rangle$} & \multicolumn{1}{c}{$\langle u_*-r \rangle$} & \multicolumn{1}{c}{$\langle M_r \rangle$} & \multicolumn{1}{c}{$M_*$} & \multicolumn{1}{c}{$M\sbr{h}$} & \multicolumn{1}{c}{$f\sbr{sat}$} & \multicolumn{1}{c}{$\chi^2$} & \multicolumn{1}{c}{d.o.f.} \\
 &  &  &  & \multicolumn{1}{c}{$10^{10} M_{\sun}$} & \multicolumn{1}{c}{$10^{11} M_{\sun}$} &  &  &  \\
215100 & 0.29 & 0.99 & -18.18 & 0.06 & $0.47 \pm 0.17$ & $0.209 \pm 0.012$ & 14.3 & 21 \\
219400 & 0.32 & 1.02 & -19.46 & 0.17 & $1.63 \pm 0.27$ & $0.179 \pm 0.010$ & 26.7 & 20 \\
80890 & 0.28 & 1.15 & -20.54 & 0.51 & $3.2 \pm 0.5$ & $0.148 \pm 0.009$ & 19.1 & 18 \\
30020 & 0.27 & 1.27 & -21.55 & 1.74 & $7.1 \pm 1.1$ & $0.114 \pm 0.007$ & 21.9 & 16 \\
3299 & 0.28 & 1.36 & -22.39 & 4.17 & $16.0 \pm 4.1$ & $0.089 \pm 0.005$ & 8.9 & 16 \\
265900 & 0.50 & 0.92 & -19.55 & 0.19 & $1.46 \pm 0.35$ & $0.176 \pm 0.010$ & 15.0 & 20 \\
179600 & 0.50 & 1.09 & -20.60 & 0.54 & $3.7 \pm 0.6$ & $0.147 \pm 0.009$ & 30.4 & 18 \\
73590 & 0.48 & 1.26 & -21.51 & 1.66 & $5.7 \pm 1.0$ & $0.115 \pm 0.007$ & 10.2 & 17 \\
16900 & 0.48 & 1.37 & -22.39 & 4.17 & $14.2 \pm 2.8$ & $0.089 \pm 0.005$ & 27.6 & 17 \\
85310 & 0.65 & 0.83 & -19.88 & 0.25 & $2.6 \pm 1.2$ & $0.168 \pm 0.010$ & 12.4 & 20 \\
239000 & 0.69 & 1.04 & -20.64 & 0.56 & $3.5 \pm 0.9$ & $0.146 \pm 0.009$ & 4.5 & 18 \\
165700 & 0.70 & 1.18 & -21.53 & 1.70 & $4.5 \pm 1.2$ & $0.115 \pm 0.007$ & 22.1 & 17 \\
44390 & 0.68 & 1.32 & -22.49 & 4.57 & $12.8 \pm 3.0$ & $0.087 \pm 0.005$ & 20.5 & 17 \\
\end{tabular}
\label{tab:bluefits}
\end{table*}

Figs. \ref{fig:delsigred} and \ref{fig:delsigblue}  show \DelSig\ as a function of projected radius for red and blue lens galaxies, respectively. The curves show the fits to the radii between $R\sbr{min}$ and 1000 kpc, based on the point mass plus NFW plus offset group halo terms.  Each panel shows a bin in redshift (increasing from bottom to top) and stellar mass (increasing left to right).  Results of the fits are also tabulated in Tables \ref{tab:redfits} and \ref{tab:bluefits}, for red and blue galaxies respectively. The fitted models are consistent with the data in all cases.

\begin{figure}
\begin{center}
\includegraphics[width=\columnwidth]{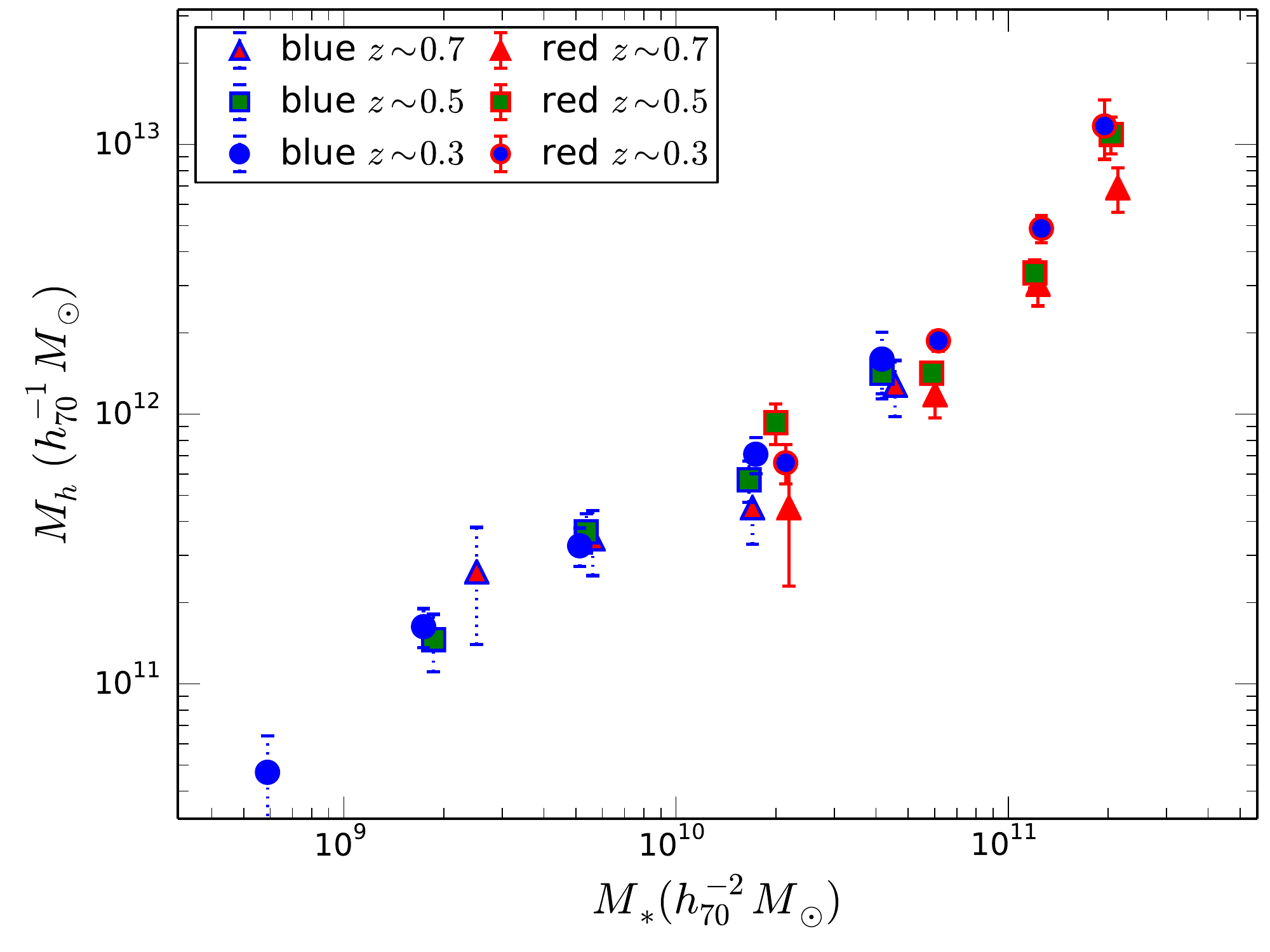}
\caption{Fitted halo mass as a function of stellar mass. Blue galaxies are shown by open symbols and red galaxies are filled symbols. The symbol type indicates the redshift of the sample, as indicated in the legend}
\label{fig:mass}
\end{center}
\end{figure}

\figref{mass} shows the fitted halo mass as a function of stellar mass \mstel\ for both red and blue galaxies at the three redshifts.  Blue galaxies dominate at low mass whereas red galaxies dominate at high masses. \figref{mass} shows that, while the halo-to-stellar mass relation of blue galaxies does not evolve as a function of redshift (within the uncertainties), that of red galaxies does. 

The slight inflection near $\mstel \sim 10^{10.5} M\sun$ indicates that the relationship between stellar mass and halo mass in nonlinear, and indeed not well described by a single power law. However the deviations from linearity are weak. Because of this, and to better appreciate the data and their uncertainties, it is more sensible to plot the stellar-to-halo-mass ratio  $f_{*} \equiv M_{*}/M\sbr{h}$ as a function of stellar mass.

\subsection{Systematics of the fit}
\label{sec:sys}

In the process of fitting the halo model, we have made several choices. It is worthwhile exploring what effect these choices have on our halo model parameters, in particular on $M_{h}$ and hence $f_{*}$.  These are (1) neglect of the two-halo term; (2) choice of the $c_{200}(M_{200})$ relation; and (3) constraining the satellite fraction to be non-evolving with a linear slope as a function of stellar mass.  We discuss each of these in more detail below.

\begin{figure*}
\begin{center}
\includegraphics[width=0.33\textwidth]{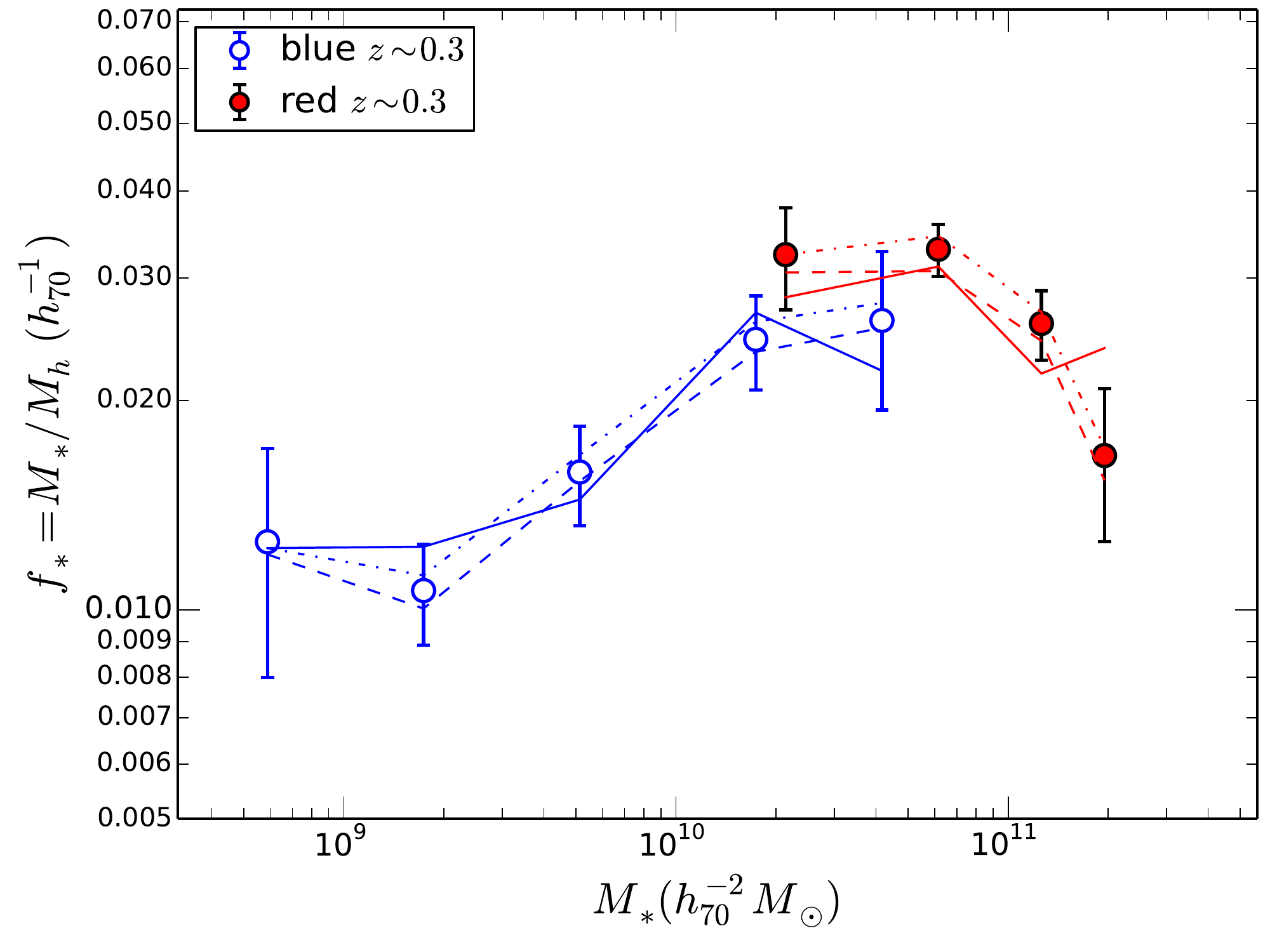}~
\includegraphics[width=0.33\textwidth]{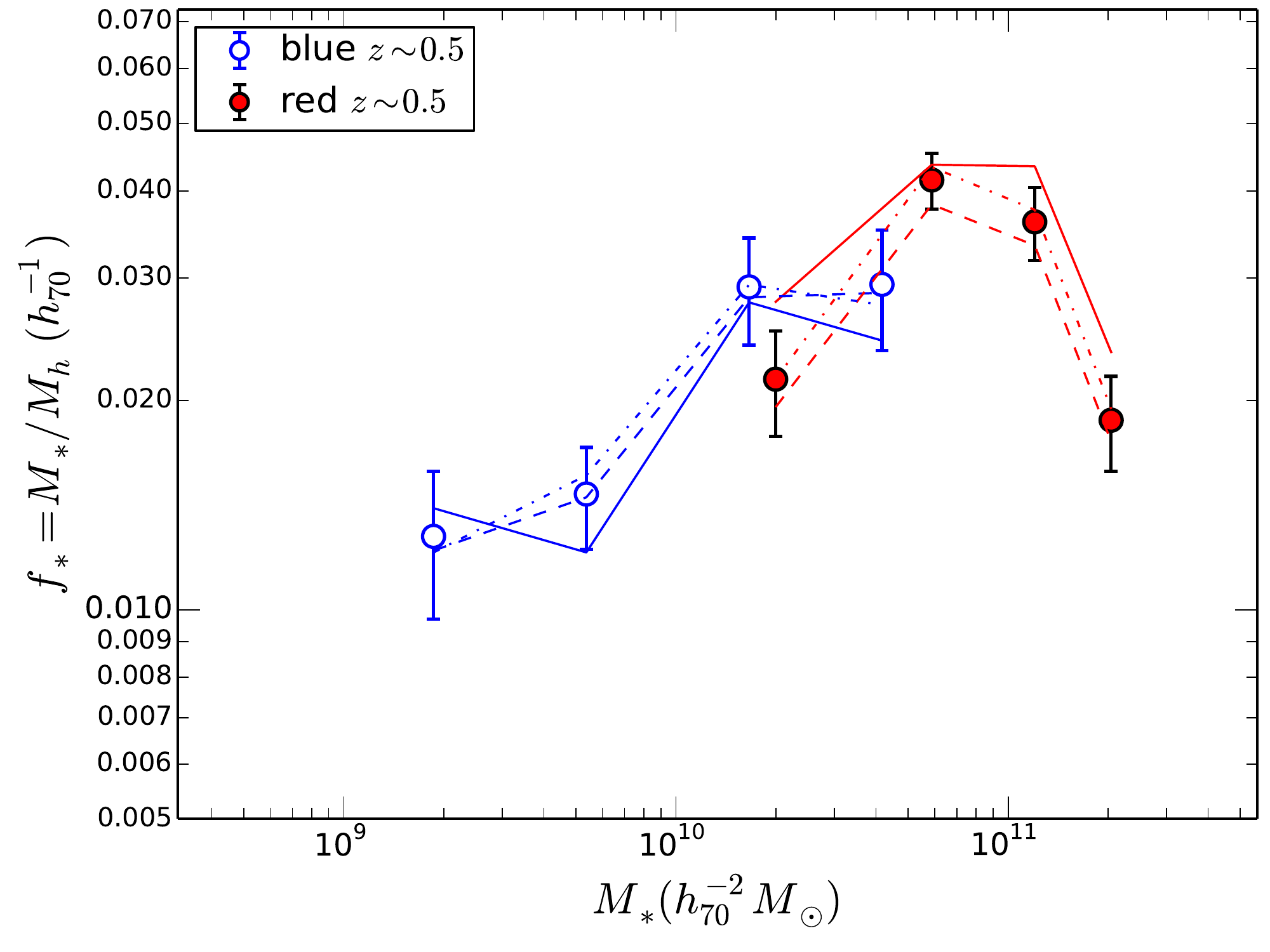}~
\includegraphics[width=0.33\textwidth]{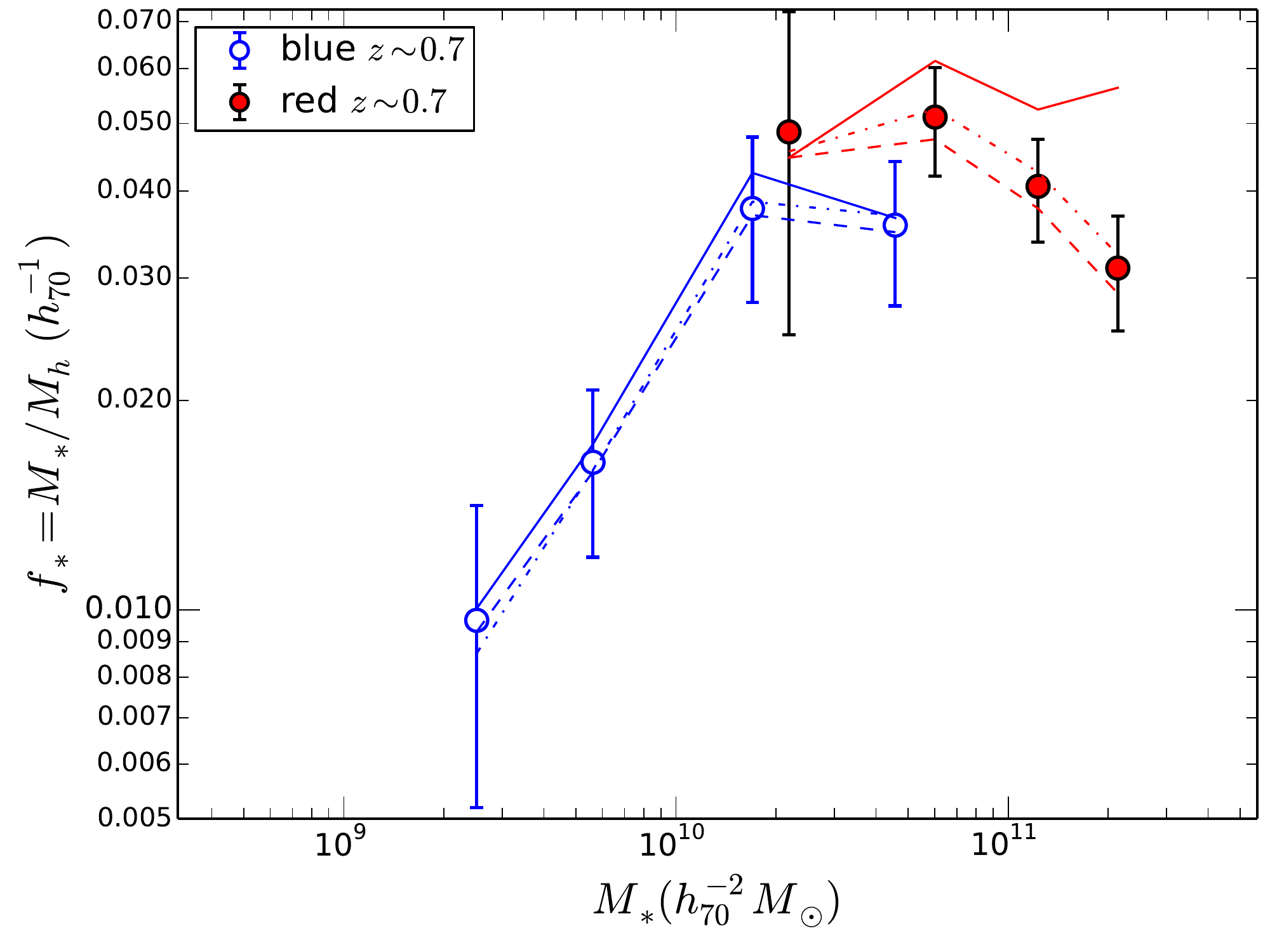}
\caption{
The SHMR $f_{*}$ as a function of stellar mass for three redshift bins ($z\sim 0.3$, $z\sim 0.5$ and $z\sim 0.7$, from left to right). The filled points show results for our default fit. The lines show the results of different assumptions in the fit, as described in more detail in the text. The dot-dash line shows the effect of including a 2-halo term, the dashed curve uses $c_{200}(M_{200})$ relation for all halos from \protect\cite{DufSchKay08}, and the solid curve shows the effect of allowing a free satellite fraction for each bin independently.}
\label{fig:sys}
\end{center}
\end{figure*}

We argued that the two-halo term is small on the scales that we are fitting ($R < 1000$ kpc). The two-halo term was calculated by V14 for the $z \sim 0.3$ redshift bin. It is a function of scale that peaks between 1000 and 2000 kpc with a peak value $\DelSig \sim 1-2$ $M_{\sun}$/pc$^{2}$, depending on the stellar mass and colour bin. This is a factor 5-10 lower than the peak of the offset-group term for red galaxies. It is perhaps only a factor of two lower than the offset-group term for blue galaxies. We have repeated our fits after having subtracted off the two-halo term estimated from V14 . We find that the satellite fractions are affected by this change: they become lower by an amount which varies from bin to bin but is at most 0.2. The halo masses, $M_{h}$ are hardly affected, however, because while offset-group term and the two-halo term compete on somewhat larger scales, and hence are partially degenerate, the halo mass is dominated by the signal on smaller scales ($R \lesssim 200$ kpc).  \figref{sys} shows the effect on the halo masses of including a two-halo term. The effect is very small compared to the random uncertainties.

The concentration-mass relation affects the fits at some level. We have adopted the relation from \cite{MunMacGot11} for relaxed haloes.  In contrast, V14 used the concentration-mass relation for all (relaxed and unrelaxed) haloes from \cite{DufSchKay08}. This makes a small difference to the fitted masses, as shown in \figref{sys}.

Note that both of these factors shift the halo masses in the same sense at all redshifts, so the relative evolution is unaffected.  

Finally, we chose to fix the satellite fraction to be the same at all redshifts for a given stellar mass and fit this with linear function in log stellar mass.  If we allow the satellite fraction to be free at all redshifts, we obtain the results shown in \figref{sys}. For blue galaxies, the effect is relatively minor as their satellite fractions are low in any case. For red galaxies, this freedom tends to increase the halo masses (lowering $f_{*}$) at $z \sim 0.3$, whereas it decreases the halo masses (raising $f_{*}$) in the two higher redshift bins.  This will increase the relative evolution between these redshift bins.

In summary, the first two systematics (two-halo term and concentration) are subdominant compared to the random errors. The effect of fitting the satellite fraction is similar to the random errors for most of the red bins (although not for the blue bins for which it is also subdominant).

\subsection{Comparison with previous galaxy-galaxy lensing results}
\label{sec:compare}

\begin{figure}
\begin{center}
\includegraphics[width=\columnwidth]{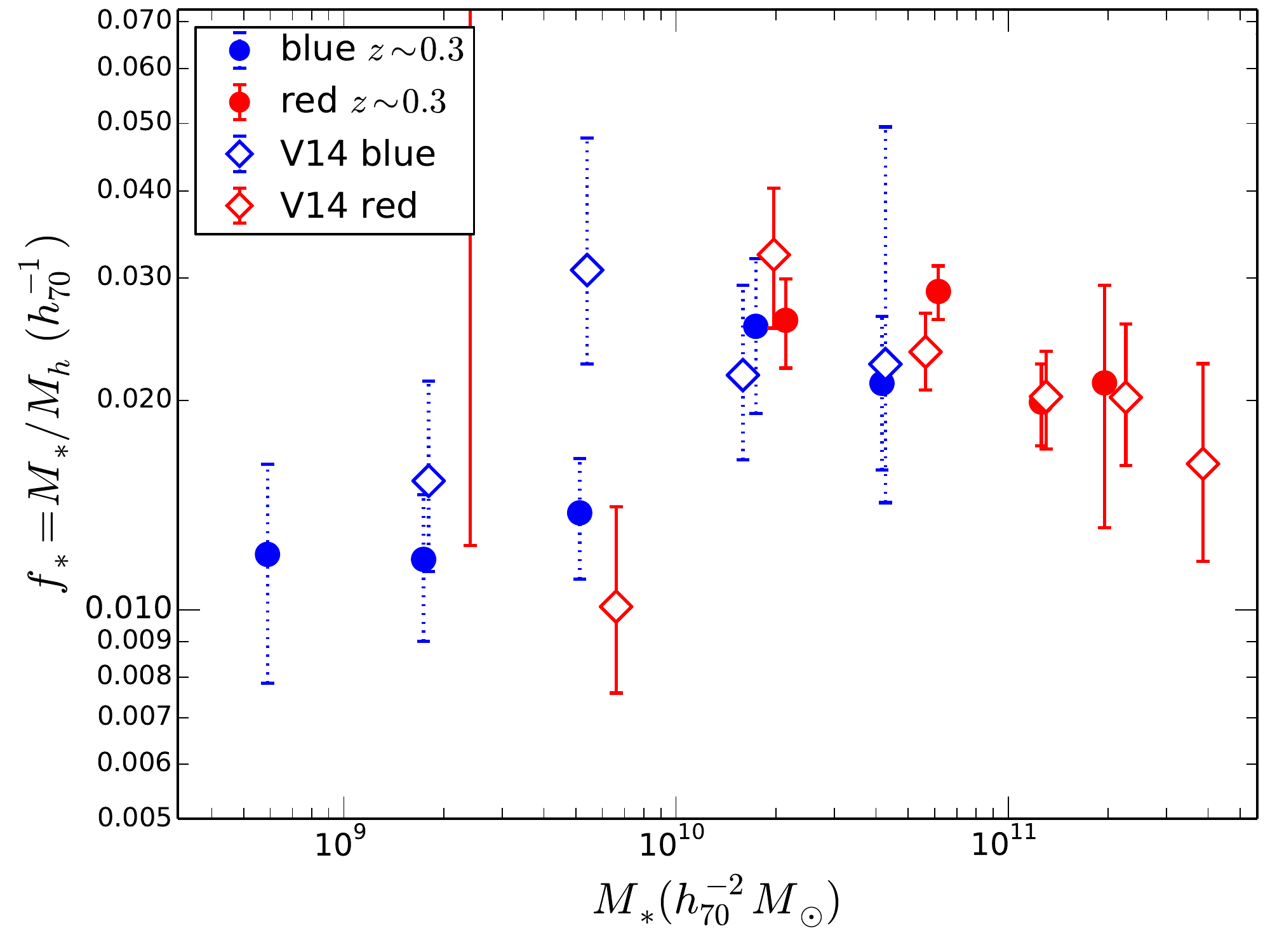}
\caption{The stellar-to-halo mass ratio (SHMR) as a function of stellar mass. The CFHTLenS results from this paper (but analysed as described in \protect\secref{compare}) at $z=0.3$ are shown by the filled circles and are compared to the CFHTLenS SHMR of V14 (open diamonds), corrected as described in the text. Red galaxies are shown in red and blue galaxies are shown by blue symbols with dotted error bars.}
\label{fig:shmr_vel}
\end{center}
\end{figure}

In \figref{shmr_vel}, we compare the results from this paper with the CFHTLenS results from V14. For this comparison, we follow the fitting method of V14 as closely as possible. In particular, first, we allow a free satellite fraction to be fit to each bin independently, and, second, we use the concentration-mass relation for all (relaxed and unrelaxed) haloes \citep[from][]{DufSchKay08}, as discussed in \secref{sys}. Nevertheless there remain important differences between the two analyses. The HOD fitting methods differ. In particular, in the V14 analysis, the shape of the offset-group term depends on the fitted halo mass, whereas in the analysis here, the shape of this term is fixed by the HOD of \cite{CouKilMcC12}. Our red/blue division is by colour, whereas that of V14 is by spectral type.  We select galaxies in a fixed bin of luminosity (and make a mean correction to \mstel), whereas V14 select by \mstel\ (and make a correction for the uncertainty in the measured \mstel).   Finally, V14 correct for the intrinsic scatter in the SHMR, so that their final data points represent the ``underlying'' SHMR. In this paper, we prefer to present the GGL data as measured, i.e. representing $\langle M_{h}|M_{*}\rangle$, and include the effect of scatter as part of the SHMR model (see \secref{shmr} below). Consequently, \figref{shmr_vel} shows the V14 points with their correction for the intrinsic scatter removed (by interpolating from their Figure B2 and the values tabulated in their Table B3). The results generally agree well. 

\begin{figure}
\begin{center}
\includegraphics[width=\columnwidth]{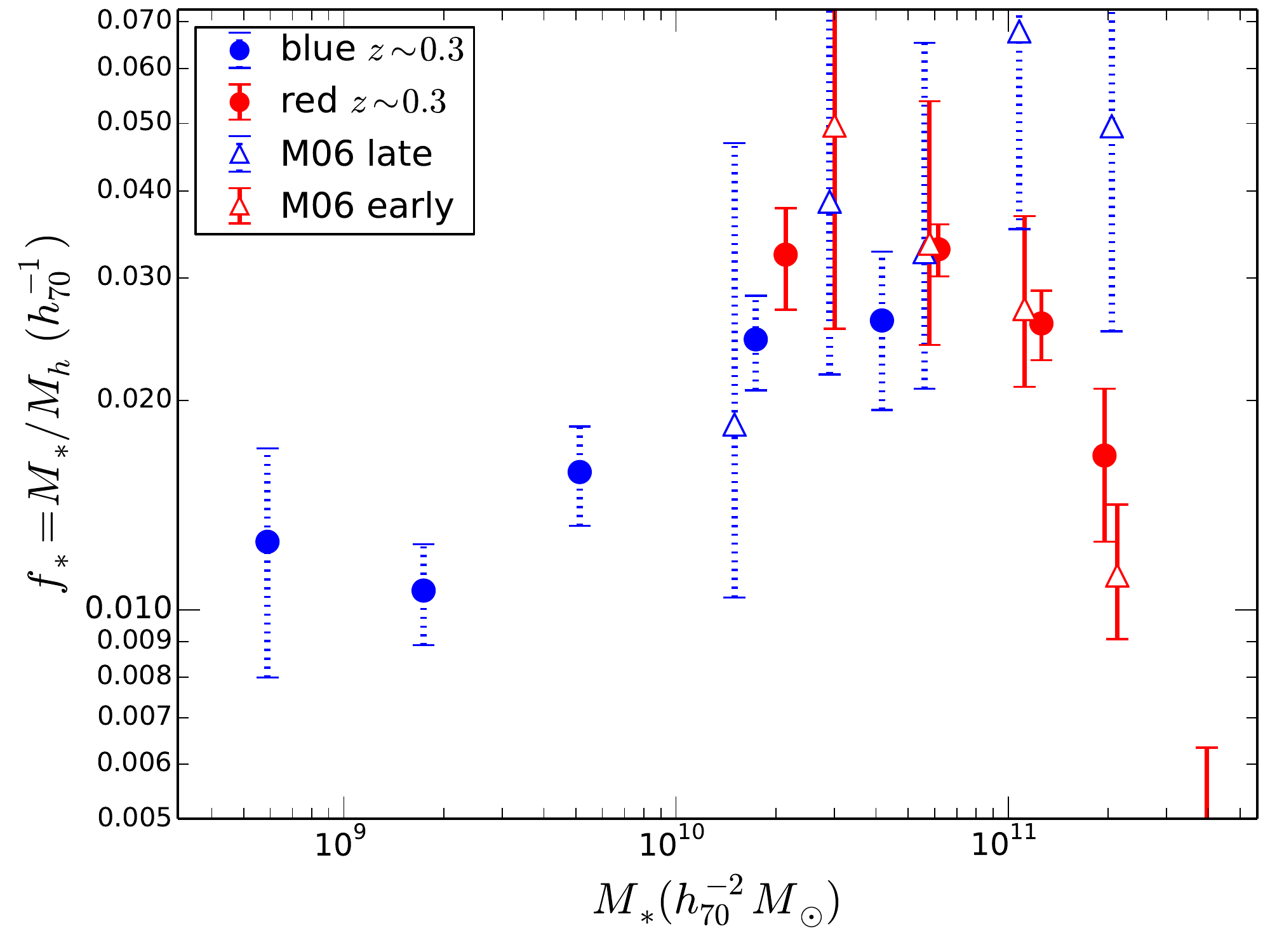}
\caption{As in \figref{shmr_vel}, but compared with the SHMR from M06 at $z = 0.1$ (after correction from $M\sbr{vir}$ to $M_{200}$) for stellar mass bins with lensing signal-to-noise greater than 2 (open triangles). Early-type (SDSS) or red (CFHTLenS) galaxies are shown in red, and late-type (SDSS) or blue (CFHTLenS) galaxies are shown by blue symbols with dotted error bars.}
\label{fig:shmr_mand}
\end{center}
\end{figure}

\figref{shmr_mand} shows our results for the SHMR, $f_{*}$, as a function of redshift, stellar mass and galaxy type, in comparison with previous results from SDSS by M06. Note that M06 selected galaxies not by colour (as is done for CFHTLenS) but by morphology.  While the SDSS data give slightly tighter constraints for rare, very massive galaxies ($\mstel \gtrsim 10^{11.5} M_{\sun}$), the CFHTLenS results are considerably tighter for less massive galaxies. Overall, the results agree very well.

\begin{figure}
\begin{center}
\includegraphics[width=\columnwidth]{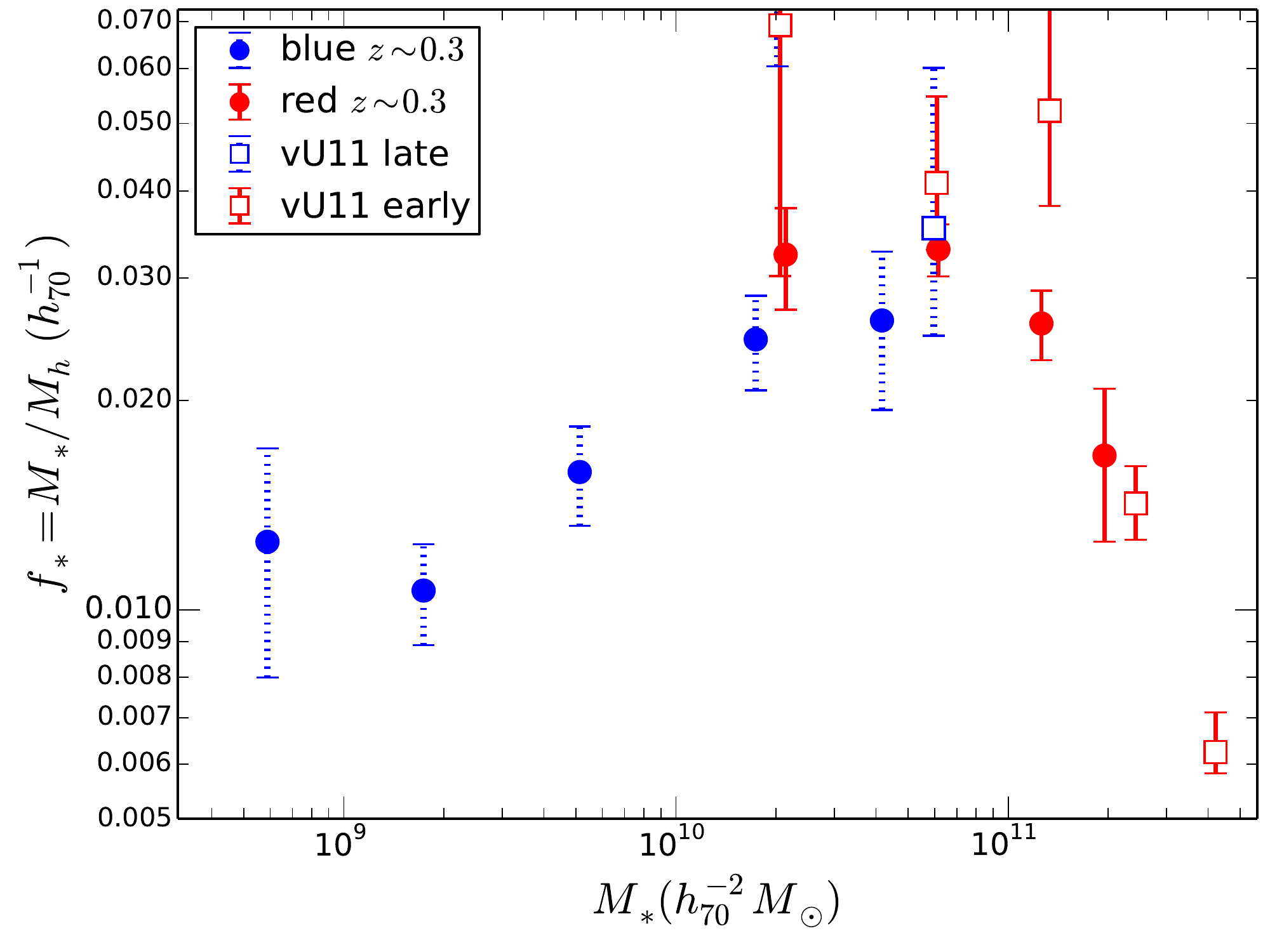}
\caption{As \figref{shmr_mand}, but CFHTLenS is compared to the SHMR of RCS2 \protect\citep{vanHoeVel11} indicated by open squares.}
\label{fig:shmr_uitert}
\end{center}
\end{figure}

\figref{shmr_uitert} shows a comparison with results from RCS2 \citep{vanHoeVel11}. These data are in the redshift range 0.08 to 0.48.  Like the SDSS, the RCS2 is wider and shallower than CFHTLS, so their constraints are tighter for very massive galaxies. 

\section{The Efficiency of Star Formation}
\label{sec:shmr}

\subsection{Parametrizing the SHMR}

In this paper, we parametrize the dependence of stellar mass on halo mass using the broken double power law relation \citep[hereafter M10]{YanMovan03, MosSomMau10}:
\begin{equation}
 f_{*}(M\sbr{h}) = 2 f_{1} \left[\left(\frac{M\sbr{h}}{M_1}\right)^{-\beta} + \left(\frac{M\sbr{h}}{M_1}\right)^{\gamma}\right]^{-1}\,.
\label{eq:moster}
\end{equation} 
With this parametrization, $M_{1}$ is a characteristic halo mass and $f_{1} = f_{*}(M_{1})$ is the stellar fraction at that mass.  The halo mass at which the SHMR peaks is $M\sbr{h,peak} = (\beta/\gamma)^{1/(\beta+\gamma)} M_{1}$. For typical values of $\beta$ and $\gamma$ found below, this yields $M\sbr{h, peak} \sim 0.87 M_{1}$. To obtain the efficiency of star formation with respect to the mean baryon density, multiply $f_{*}$ by $\Omega\sbr{m}/\Omega\sbr{baryon} = 6.36$ \citep[][CMB plus BAO]{Planck13-16}. 
Our fits do not constrain very strongly the high-mass SHMR slope parameter, $\gamma$, although they disfavour the M10 value of 0.6, at the $\sim 2\sigma$ level, and prefer higher values. We therefore fix $\gamma = 0.8$ in the following.

We will fit the SHMR parameters allowing $f_{1}$ and $\log_{10}(M_{1})$ to evolve linearly with redshift. The CFHTLenS lenses are centred at approximately $z = 0.5$, so we Taylor-series expand around this redshift and adopt
\begin{equation}
f_{1}(z) = f_{0.5} + (z - 0.5) f_{z}
\end{equation}
and
\begin{equation}
\log_{10}(M_{1})(z) = \log_{10} M_{0.5} + (z - 0.5) M_{z}\,.
\end{equation}
Note that even if $M_{z}$ is consistent with zero so that there is no dependence of $f_{*}$ as a function of halo mass (i.e.\ no halo-mass downsizing), it is still possible to have downsizing in $f_{*}$  as a function of stellar mass if $f_{z}$ is significantly different from zero.

Note that the functional form that we have adopted, \equref{moster}, yields the mean SHMR as a function of halo mass.  One could plot the SHMR as a function of halo mass $f_{*}(M\sbr{h})$, but in practice this is complicated because the halo mass is the measured quantity with largest uncertainties. This would complicate the plots and uncertainties because halo mass would appear in both the independent and dependent variables.   In contrast, the \emph{observational} uncertainty in the stellar mass is negligible compared to that in the halo mass, and so it is more sensible to treat stellar mass as the independent variable. 

The complication with using stellar mass as the independent variable is that we expect individual galaxies to scatter around this mean relation (M10).   In this paper, we will adopt a scatter of 0.15 dex \citep[following M10 and][]{BehConWec10}, independent of mass, around $f_{*}(M\sbr{h})$.
To obtain $\langle M_{h}|M_{*} \rangle$, we integrate over the abundance of halo masses and this lognormal scatter:
\begin{equation}
\langle M\sbr{h} | M\sbr{*} \rangle = 
\frac{\int_{0}^{\infty} M\sbr{h} \, P(M\sbr{*, pred} | M_{*})) \, N(M\sbr{h}) dM\sbr{h} }{\int_{0}^{\infty} P(M\sbr{*, pred} | M_{*})) \, N(M\sbr{h}) dM\sbr{h} }
\label{eq:mstelscat}
\end{equation}
where $M\sbr{*, pred} = M_{h} \times f_{*}(M\sbr{h})$  and $P$ is the lognormal distribution of the stellar mass as a function of the predicted stellar mass and $N$ is the abundance of halos of mass $M\sbr{h}$ \citep{MurPowRob13}.  For readers who want a simple parametric expression for the estimated halo mass as a function of stellar mass, in Appendix C we present an alternative parametric fit.

\subsection {Evolution of red and blue galaxies}
\label{sec:redblue}

\begin{figure*}
\begin{center}
\includegraphics[width=0.49\textwidth]{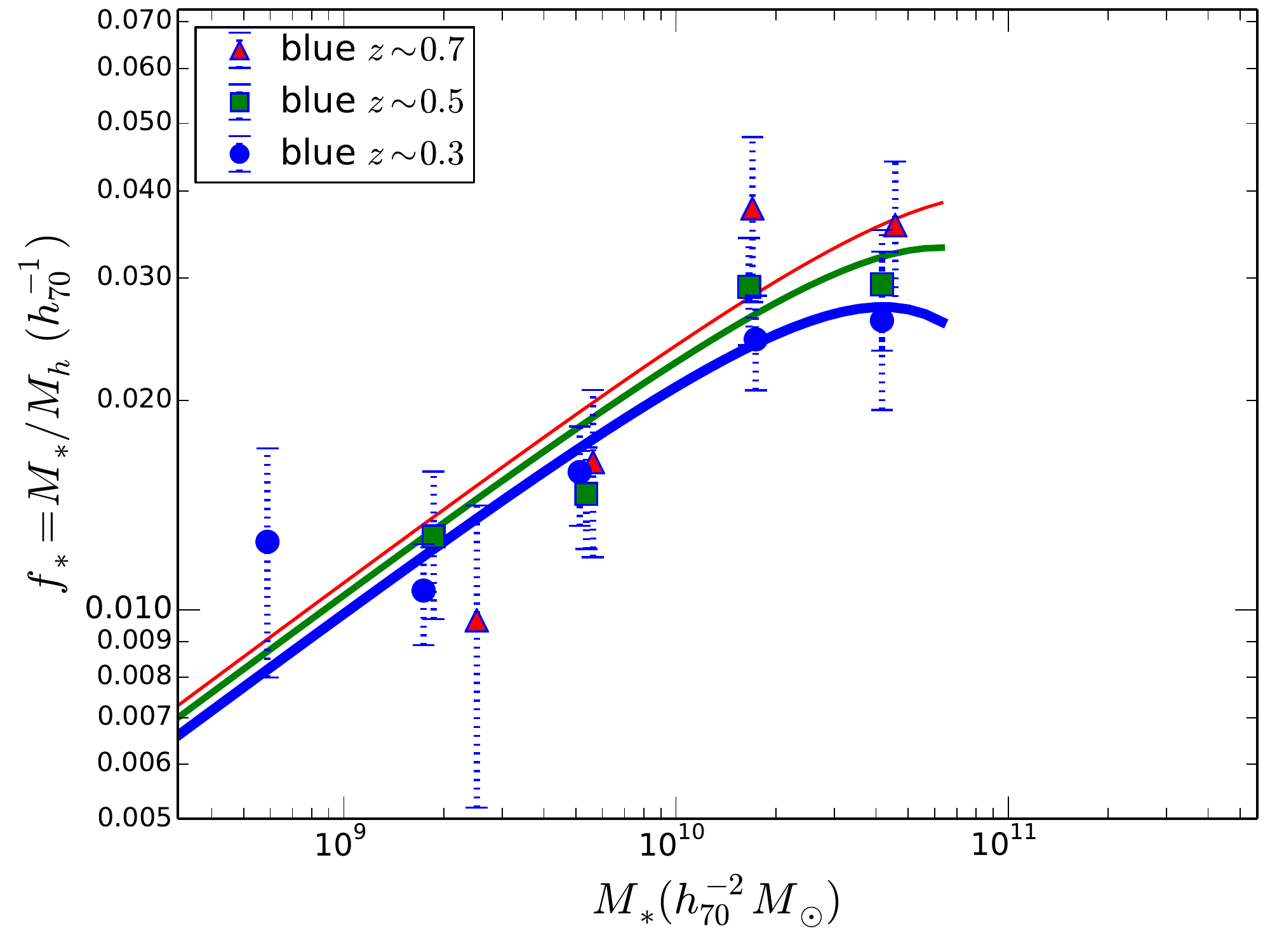}~
\includegraphics[width=0.49\textwidth]{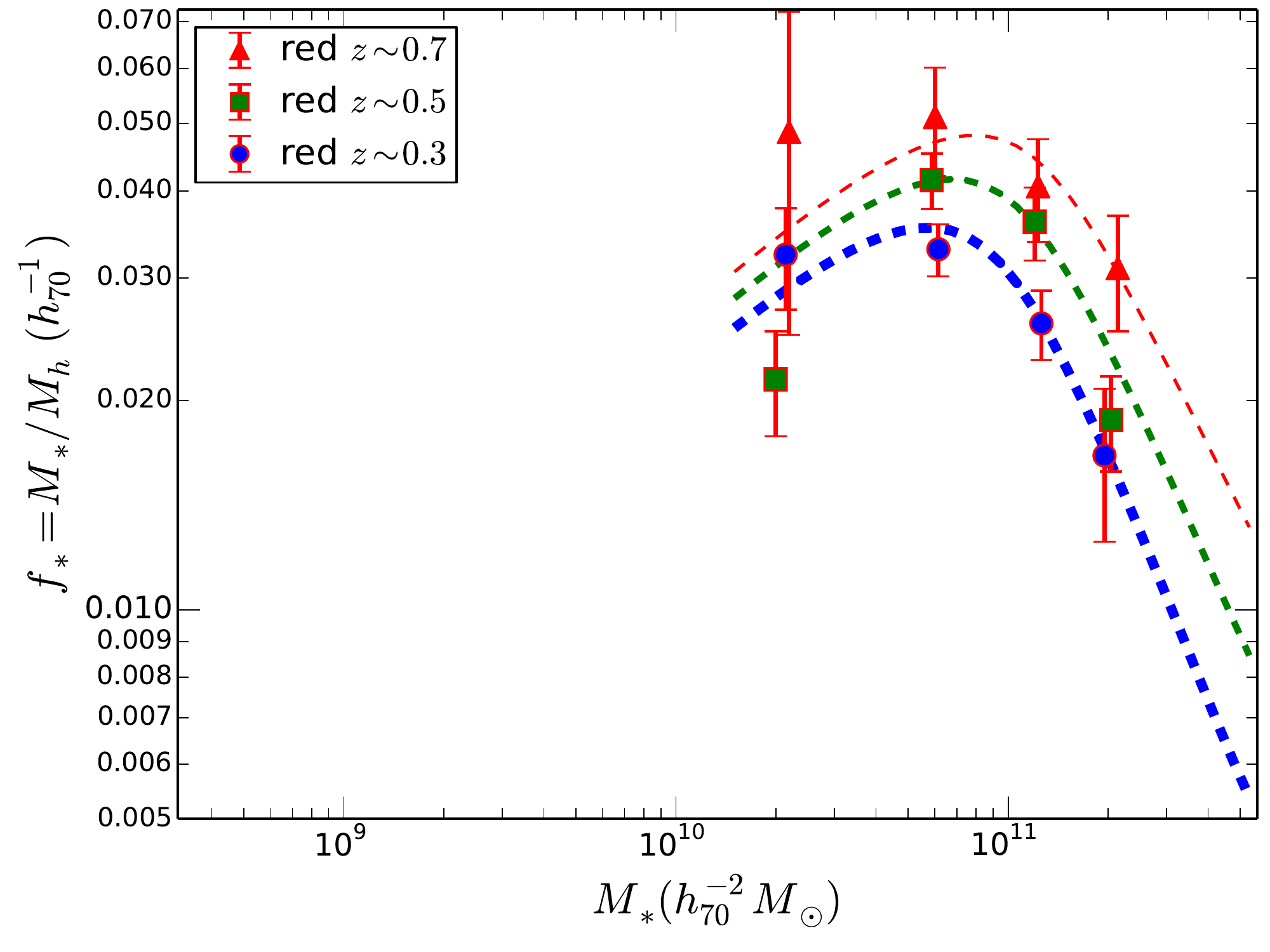}
\caption{Left: The stellar-to-halo mass fraction (SHMR) as a function of stellar mass for blue galaxies. Data from three different redshift ranges ($z=0.3$, $0.5$ and $0.7$ shown in blue circles, green squares and red triangles, respectively). The lines are M10 double power law fits to the evolution (``Default'' fit from \tabref{shmrblue}), with thickest lines indicating the fit to the lowest redshift bin. Right: same for red galaxies, based on the ``Default'' fit from \tabref{shmrred}. The evolution is clear: the peak $f_{*}$ of the red galaxies decreases and shifts to lower stellar masses at later epochs.}
\label{fig:shmr_redblue}
\end{center}
\end{figure*}

We will first discuss the red and blue populations separately.  Blue galaxies dominate at low stellar masses whereas red galaxies dominate at high stellar masses. Consequently, there are only two mass bins (in the range $10^{10}M_{\sun} \lesssim M_{*} \lesssim 10^{11} M_{\sun}$) where there are sufficient numbers of both red and blue galaxies in the bin to obtain a halo mass measurement for both the blue and red populations. The independent fits of the M10 double power law given by \equref{moster} to red and blue populations, both as a function of redshift, are plotted in \figref{shmr_redblue}.

\makeatletter{}\begin{table*}
\newcolumntype{d}[1]{D{.}{.}{#1} }
\caption{Double power-law (M10) fits to the SHMR of blue galaxies}
\begin{tabular}{lcccccccd{-1}r}
\multicolumn{1}{c}{label} & \multicolumn{1}{c}{$f_{0.5}$} & \multicolumn{1}{c}{$f_z$} & \multicolumn{1}{c}{$\log(M_{0.5})$} & \multicolumn{1}{c}{$M_z$} & \multicolumn{1}{c}{$\beta$} & \multicolumn{1}{c}{$\gamma$} & \multicolumn{1}{c}{$\delta\sbr{blue}$} & \multicolumn{1}{c}{$\chi^2$} & \multicolumn{1}{c}{d.o.f.} \\
\hline
Default & $0.034 \pm 0.010$ & $0.03 \pm 0.05$ & $12.5 \pm 0.6$ & $0.4 \pm 1.5$ & $0.55 \pm 0.22$ & 0.8 & 0.0 & 7.7 & 8 \\
No Evolution & $0.039 \pm 0.035$ & 0.00 & $12.7 \pm 1.3$ & 0.0 & $0.52 \pm 0.22$ & 0.8 & 0.0 & 9.3 & 10 \\
Single Power Law & $0.044 \pm 0.012$ & $0.028 \pm 0.021$ & 13.0 & 0.0 & $0.45 \pm 0.08$ & 0.8 & 0.0 & 7.9 & 10 \\
\label{tab:shmrblue}
\end{tabular}
\end{table*}
 
\makeatletter{}\begin{table*}
\newcolumntype{d}[1]{D{.}{.}{#1} }
\caption{Double power-law (M10) fits to the SHMR of red galaxies}
\begin{tabular}{lcccccccd{-1}r}
\multicolumn{1}{c}{label} & \multicolumn{1}{c}{$f_{0.5}$} & \multicolumn{1}{c}{$f_z$} & \multicolumn{1}{c}{$\log(M_{0.5})$} & \multicolumn{1}{c}{$M_z$} & \multicolumn{1}{c}{$\beta$} & \multicolumn{1}{c}{$\gamma$} & \multicolumn{1}{c}{$\delta\sbr{blue}$} & \multicolumn{1}{c}{$\chi^2$} & \multicolumn{1}{c}{d.o.f.} \\
\hline
Default & $0.045 \pm 0.005$ & $0.034 \pm 0.012$ & $12.27 \pm 0.13$ & $0.00 \pm 0.08$ & $0.8 \pm 0.5$ & 0.8 & 0.0 & 6.8 & 7 \\
No char.\ halo mass evol. & $0.045 \pm 0.004$ & $0.034 \pm 0.009$ & $12.27 \pm 0.11$ & 0.0 & $0.8 \pm 0.4$ & 0.8 & 0.0 & 6.8 & 8 \\
No peak $f_*$ evol. & $0.042 \pm 0.007$ & 0.00 & $12.33 \pm 0.19$ & $0.65 \pm 0.29$ & $0.7 \pm 0.5$ & 0.8 & 0.0 & 13.1 & 8 \\
No Evolution & $0.043 \pm 0.007$ & 0.00 & $12.28 \pm 0.19$ & 0.0 & $0.9 \pm 0.8$ & 0.8 & 0.0 & 22.0 & 9 \\
\label{tab:shmrred}
\end{tabular}
\end{table*}

The SHMR fits to blue galaxies are given in \tabref{shmrblue}. The SHMR of blue galaxies does not evolve significantly as a function of redshift: the redshift dependence of $f_{*}$ at fixed mass is only $f_z = 0.03 \pm 0.05$, obviously consistent with zero. The best fitting M10 function has a low-mass-slope that is $\beta = 0.55\pm0.22$. In this case, the errors on $\beta$ are large because of partial degeneracies between $\beta$ and the break mass $M_{0.5}$. A model with no evolution is a good fit. Its $\chi^{2}$ is larger by only 1.6 with 2 more degrees of freedom. If, instead, we fit a single power law SHMR to the blue galaxies (last row of \tabref{shmrblue}), the uncertainties on the slope are tighter: $\beta = 0.45\pm0.08$, as are the constraints on the evolution: $f_{z} = 0.028\pm0.021$. This is also a better fit than the M10 function, with a $\chi^{2}$ value larger by only 0.2, but with two more degrees of freedom. Thus the SHMR of blue galaxies is well-described by a non-evolving (single) power law.

In contrast to the blue galaxy population, for the red galaxies a model with no evolution (both $f_{z} = 0$ and $M_{z} = 0$, last row in \tabref{shmrred}) is a poor fit ($\chi^2 = 22$ for 9 d.o.f.). The model labelled ``Default'' in \tabref{shmrred} has both evolution in the SHMR normalization $(f_{z} \ne 0)$ and evolution in the characteristic halo mass $(M_{z} \ne 0)$. However, it is not a statistically significant improvement over the simpler model with evolution in the normalization but no mass downsizing ($M_{z} = 0$, labelled ``No char.\ halo mass evol.''). Thus while the downsizing term in halo mass is not significant, the evolution of the normalization is highly significant: $f_z = 0.034\pm0.009$.  Compared to the ``No char.\ halo mass evol.'' model, the no evolution model has $\Delta \chi^{2} = 15.2$ for 1 more degree of freedom, and so is formally disfavoured at the 99.99\% confidence level (CL), or $3.9 \sigma$. 

\subsection{Fits to all galaxies}
\label{sec:all}

\begin{figure}
\begin{center}
\includegraphics[width=\columnwidth]{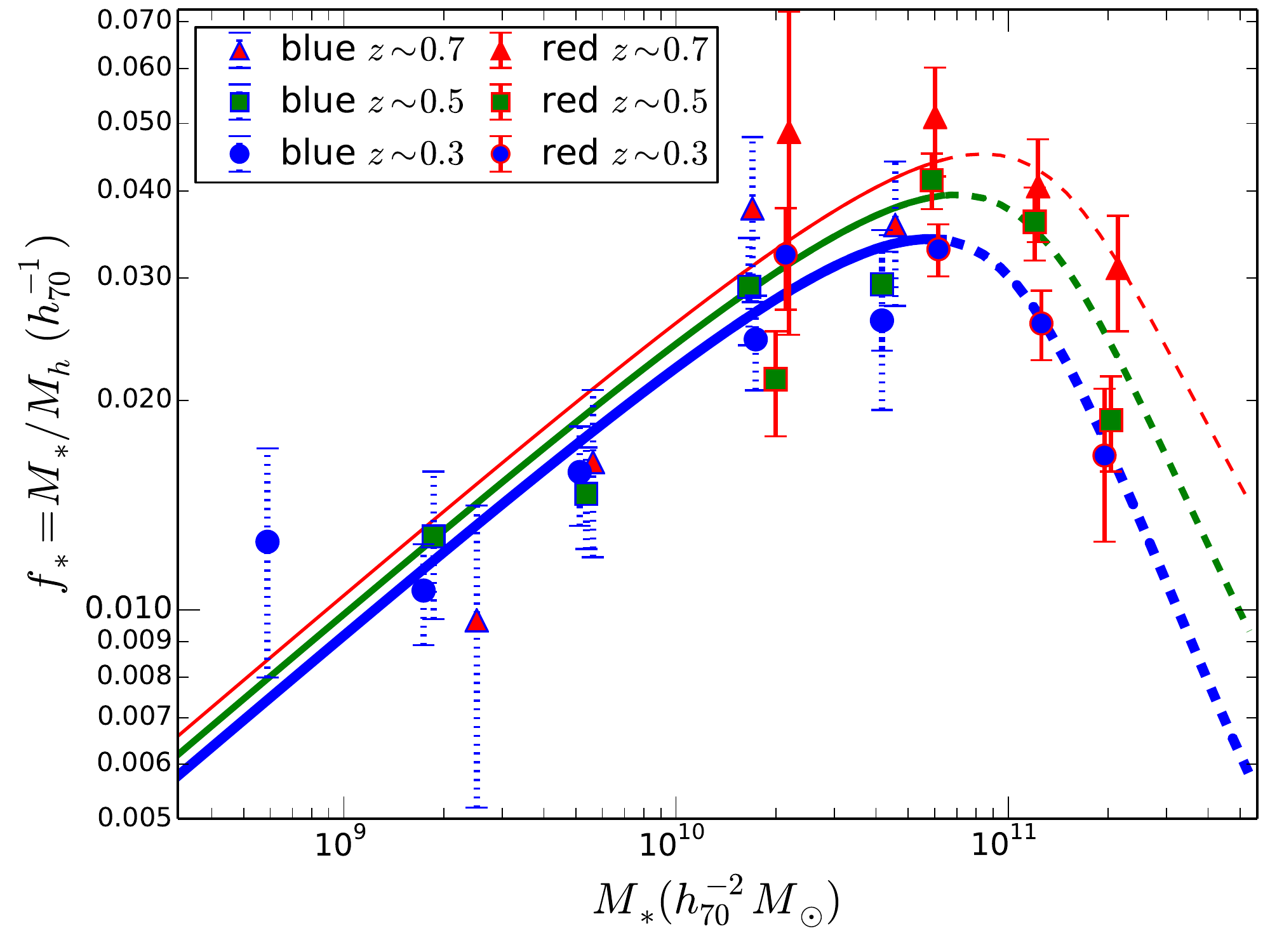}
\caption{As in \figref{shmr_redblue}, but with a single M10 model fit to both red and blue galaxies simultaneously.}
\label{fig:shmr_all}
\end{center}
\end{figure}

Some analyses of the galaxy SHMR do not explicitly distinguish between red and blue galaxies, and so it is interesting to consider the SHMR for all galaxies, independent of their colour. We have fit the red and blue data simultaneously, so each sample is effectively weighted by its inverse square errors. The results of the fits are given in \tabref{shmrall}, and shown in \figref{shmr_all}. 
The red galaxies dominate the peak of the SHMR, so the downsizing effect discussed above is also present here: the peak of the SHMR shifts to lower stellar masses at lower redshifts. A model with no evolution (i.e.\ a fixed SHMR independent of redshift) has a high $\chi^{2} = 33.3$ for 22 degrees of freedom.  
As with the red galaxies, the model labelled ``Default'' that has both evolution in the SHMR normalization and evolution in the characteristic halo mass is a better fit than the ``No evolution'' model.  However, it is not a statistically significant improvement over the simpler model with evolution in the normalization but no mass downsizing (labelled ``No char.\ halo mass evol.''). In the latter model, the redshift dependence of the normalization, $f_{z} = 0.031\pm0.008$, is highly significant. Compared to the ``No char.\ halo mass evol.'' model, the no evolution model has $\Delta \chi^{2} = 15.0$ for 1 more degree of freedom, and so is formally disfavoured at the 99.99\% CL (equivalent to $\sim 3.9 \sigma$ significance).
For the default fit, the peak $f_{*}$ drops from 
$0.045 \pm 0.003$ at $z = 0.7$ to  
$0.040 \pm 0.002$ at $z = 0.5$ and  
$0.034 \pm 0.002$ at $z = 0.3$.   
The \emph{stellar} mass of the peak ``downsizes'' from 
$\left(7.8 \pm 1.0\right) \times 10^{10} M_{\sun}$ to 
$\left(6.7 \pm 0.5\right) \times 10^{10} M_{\sun}$  to 
$\left(5.6 \pm 0.4\right) \times 10^{10} M_{\sun}$  at the same redshifts.
As with red galaxies, downsizing in \emph{halo} mass is not significantly different from zero: $M_{z} = 0.09\pm0.24$.  This suggests a picture where the halo mass of $10^{12.25} M_{\sun}$ is a time-independent peak mass above which the ratio of stellar-to-halo mass declines.

We saw above that red and blue galaxies may follow slightly different SHM relations.  There have been suggestions of small differences in the SHMR of red and blue galaxies. Using satellite kinematics, \cite{MorvanCac11} found no significant difference between red and blue centrals for stellar masses less than $10^{10.5} M_{\sun}$, but for stellar masses greater than that value, blue galaxies had slightly lower halo masses than red galaxies.  \cite{WanWhi12} find that isolated red galaxies have more satellites (possibly a proxy for halo mass) per unit stellar mass than blue galaxies.  However, \citet[hereafter T13]{TinLeaBun13} find that passive (red) galaxies have lower masses than active (blue) galaxies at given \mstel\ (see also \figref{Leauthaud} discussed in \secref{compareshmr} below). For the CFHTLenS data, there appears to be no difference between red and blue populations at $\mstel \sim 10^{10.3} M_{\sun}$, but the SHMRs of blue galaxies are lower at $\mstel \sim 10^{10.7} M_{\sun}$. We have fit the blue and red populations simultaneously, allowing for an offset term for the halo masses of blue galaxies: $M_{200}(\text{blue}) = (1+\delta\sbr{blue}) M_{200}(\text{red})$. We find hints of a difference between blue and red galaxies, with blue galaxies having slightly more massive haloes at fixed stellar mass (in agreement with T13), but the offset is not statistically significant: specifically, $\delta\sbr{blue} = 0.22\pm0.13$.  

\makeatletter{}\begin{table*}
\newcolumntype{d}[1]{D{.}{.}{#1} }
\caption{Double power-law (M10) fits to the SHMR of all galaxies}
\begin{tabular}{lcccccccd{-1}r}
\multicolumn{1}{c}{label} & \multicolumn{1}{c}{$f_{0.5}$} & \multicolumn{1}{c}{$f_z$} & \multicolumn{1}{c}{$\log(M_{0.5})$} & \multicolumn{1}{c}{$M_z$} & \multicolumn{1}{c}{$\beta$} & \multicolumn{1}{c}{$\gamma$} & \multicolumn{1}{c}{$\delta\sbr{blue}$} & \multicolumn{1}{c}{$\chi^2$} & \multicolumn{1}{c}{d.o.f.} \\
\hline
Default & $0.0414 \pm 0.0024$ & $0.029 \pm 0.009$ & $12.36 \pm 0.07$ & $0.09 \pm 0.24$ & $0.69 \pm 0.09$ & 0.8 & 0.0 & 18.1 & 20 \\
No char.\ halo mass evol. & $0.0415 \pm 0.0024$ & $0.031 \pm 0.008$ & $12.36 \pm 0.07$ & 0.0 & $0.68 \pm 0.09$ & 0.8 & 0.0 & 18.3 & 21 \\
No peak $f_*$ evol. & $0.0411 \pm 0.0028$ & 0.00 & $12.33 \pm 0.08$ & $0.49 \pm 0.23$ & $0.75 \pm 0.12$ & 0.8 & 0.0 & 27.9 & 21 \\
No Evolution & $0.0399 \pm 0.0029$ & 0.00 & $12.38 \pm 0.09$ & 0.0 & $0.69 \pm 0.12$ & 0.8 & 0.0 & 33.3 & 22 \\
With Offset & $0.0415 \pm 0.0025$ & $0.030 \pm 0.009$ & $12.36 \pm 0.07$ & $0.06 \pm 0.24$ & $0.56 \pm 0.10$ & 0.8 & $0.22 \pm 0.13$ & 15.3 & 19 \\
\label{tab:shmrall}
\end{tabular}
\end{table*}

\subsection{Comparison with other results}
\label{sec:compareshmr}

\begin{figure}
\begin{center}
\includegraphics[width=\columnwidth]{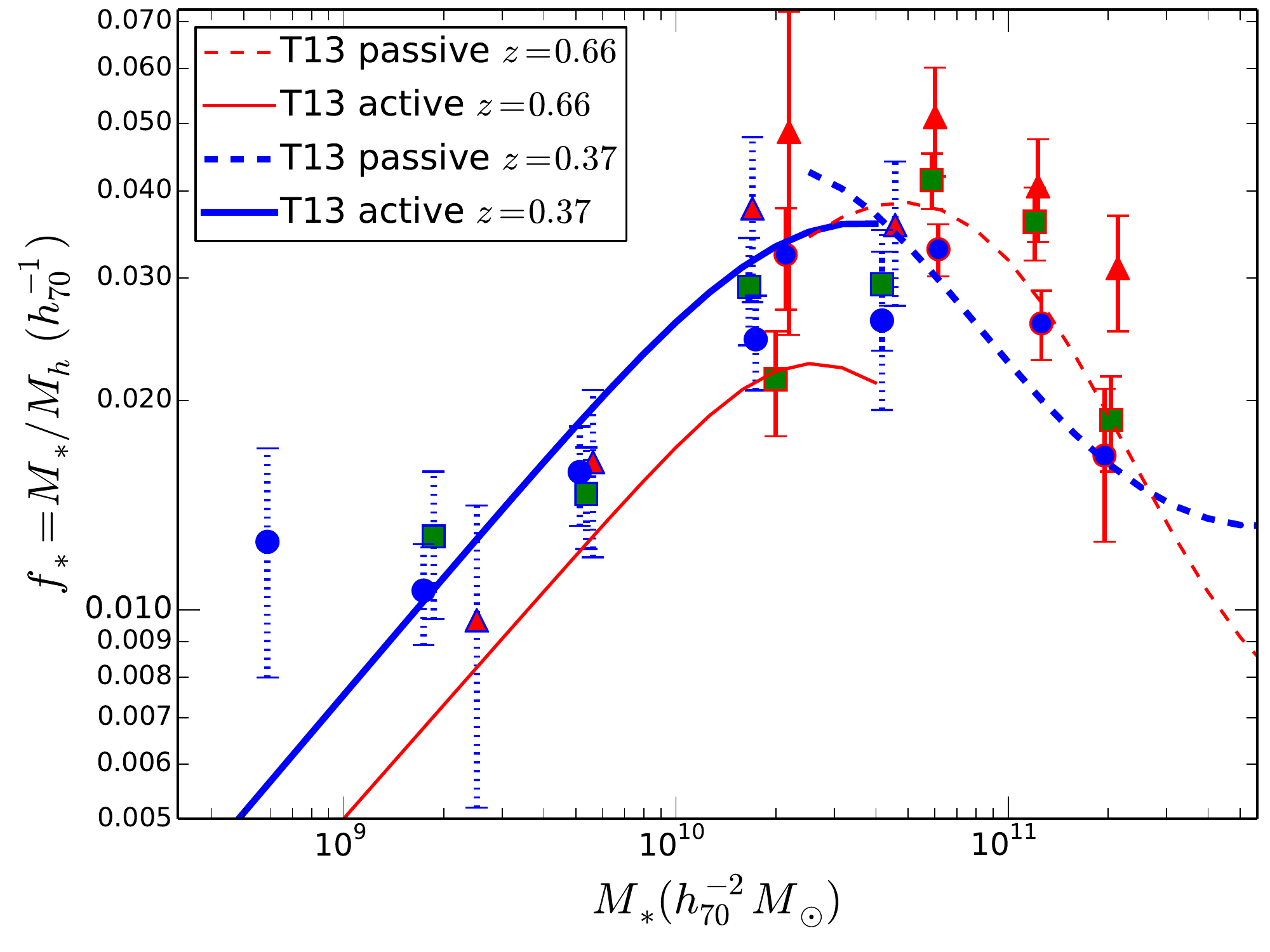}
\caption{The SHMR for the CFHTLenS sample (symbols as in \protect\figref{shmr_redblue}) compared to the model fits of T13, which are based on a combination of weak lensing, clustering and abundance matching. Compare the thin red $z=0.66$ dashed and solid lines (for passive and active galaxies, respectively) to the CFHTLenS $z\sim 0.7$ galaxies in red triangles. At lower redshift, compare the $z=0.37$ model is closest in redshift to the CFHTLenS $z\sim 0.3$ galaxies shown as blue circles.}
\label{fig:Leauthaud}
\end{center}
\end{figure}

In \secref{compare}, we compared our results to other weak lensing results. Here we compare our SHMR derived solely from weak lensing with other methods. T13, extending previous work of L12, performed a joint analysis of weak lensing, clustering and abundance for active and passive galaxies in the COSMOS field. Their parametric fits, converted to $f_{*} = M_{*}/\langle M\sbr{h}|M_{*}\rangle$, and correcting for their different definition of halo mass (200 times background density), are compared with our weak lensing data in \figref{Leauthaud}.  The approximate peak location and peak heights are comparable, and their low-mass and high-mass slopes are similar for both T13 and CFHTLenS. 

We noted in \secref{sys}, that there were small systematics associated with the fitting method. In addition, one difference in modelling that can account for some of the discrepancy between T13 and our results is the treatment of partially-stripped satellite subhalos. This issue is discussed in greater detail in Appendix \ref{sec:subhalo}. There are also systematic uncertainties (at the level of 0.2 dex) associated with the stellar mass estimates, both for our sample and that of T13. 

On the other hand, the evolution of the SHMR is a \emph{differential} measurement, and so we expect it to be more robust to systematic uncertainties than the absolute value of the SHMR. The fits of T13 indicate that there is evolution in the low-mass blue galaxy SHMR (compare the $z=0.37$ SHMR with the $z=0.66$ SHMR in \figref{Leauthaud}) that we do not observe. L12 also suggested that evolution of the SHMR evolution could be described by a model in which the peak star formation efficiency did not depend on redshift, but where the peak (``pivot'') halo mass decreased with time. In our fits (Table \ref{tab:shmrall}), such models are labelled ``No peak $f_*$ evol.'' and are generally disfavoured at the $\sim 2\sigma$ level in comparison to models in which the normalization (and hence the peak $f_{*}$) depends on redshift. While we do find downsizing in the peak \emph{stellar} mass, we do not find significant evidence for downsizing in the location of the peak \emph{halo} mass from our CFHTLenS data. 

\begin{figure}
\begin{center}
\includegraphics[width=\columnwidth]{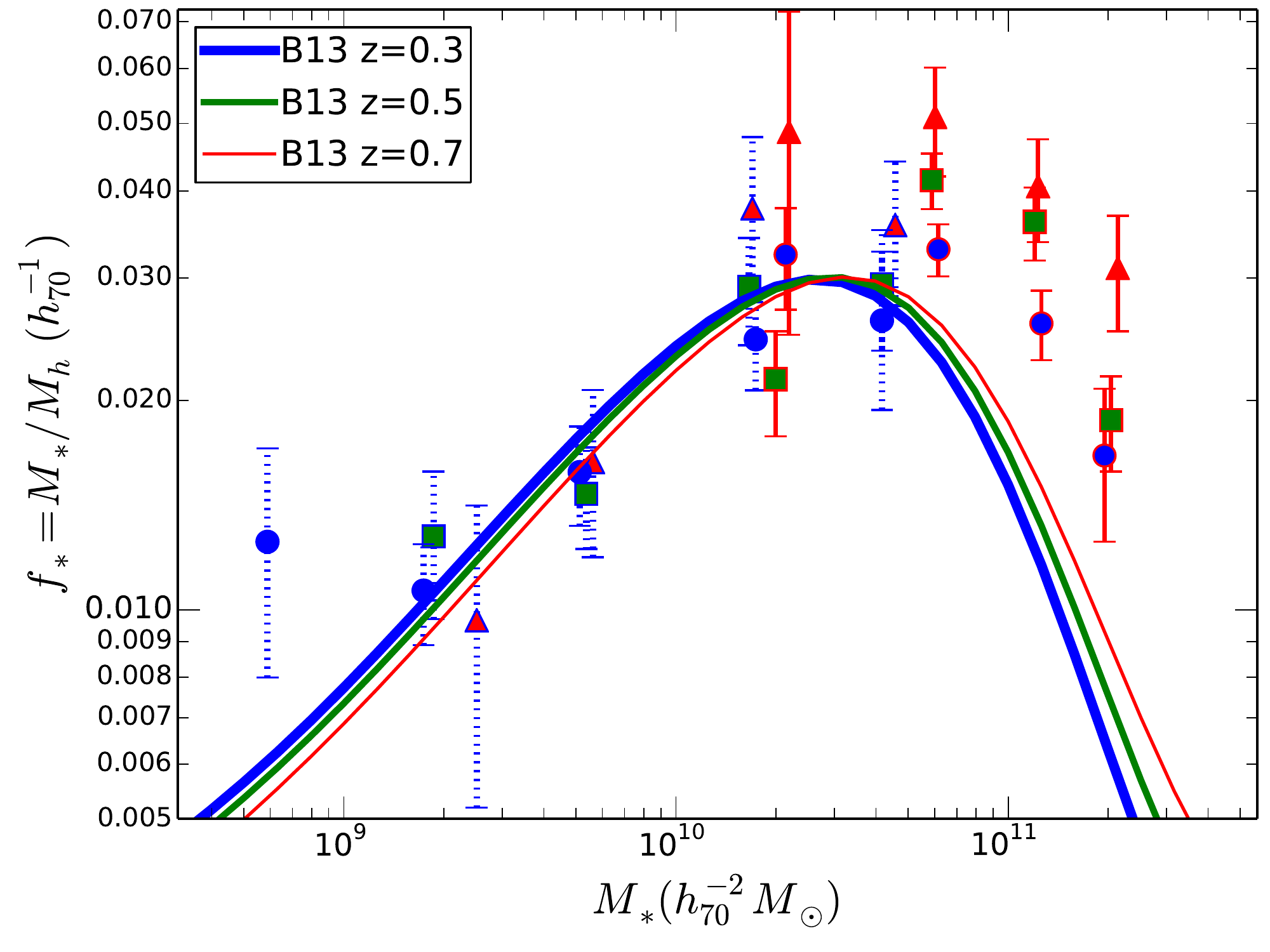}
\caption{The SHMR for the CFHTLenS sample (symbols as in \figref{shmr_redblue}) compared to the abundance matching fits of \protect\cite{BehWecCon13} as indicated in the legend.}
\label{fig:Behroozi}
\end{center}
\end{figure}

\cite{BehWecCon13} compiled data on the stellar mass function, the specific star formation rate and the cosmic star formation rate from a number of sources and used abundance matching to fit the SHMR and its evolution over a wide range of redshifts.  In \cite{BehConWec10}, the systematic uncertainties in abundance matching are discussed in detail.  The largest of these, the uncertainty in stellar mass estimates, leads to systematic uncertainties in the SHMR of order 0.25 dex. 
The fits of \cite{BehWecCon13}  are overlaid on the CFHTLenS data in \figref{Behroozi}. Although their models underpredict the CFHTLenS $f_{*}$ for red galaxies (particularly at higher redshift), the difference is within the 0.25 dex uncertainty. The overall shape of the SHMR and its dependence on redshift are similar to those observed in the CFHTLenS data. In particular, the low-mass slope and the weak evolution of the SHMR of faint blue galaxies are consistent with the shallow $\beta$ and its lack of evolution. Although offset from the GGL data at masses greater than the peak mass, their model predicts evolutionary \emph{trends} at high mass that are consistent with what we observe, albeit with less evolution than we find.  Specifically, at  fixed stellar mass $M_{*}= 10^{11}M_{\sun}$, the evolution predicted by their model is $f_{*}(z=0.7)/f_{*}(z=0.3) = 1.23$, whereas we find  $f_{*}(z=0.7)/f_{*}(z=0.3) = 1.36\pm0.09$.

\subsection{Faint blue dwarfs}
\label{sec:faintbluedwarfs}

The power of the CFHTLenS sample allows us to measure the DM halo masses of faint blue galaxies,  with mean luminosities $M_{r} \sim -18$ or, equivalently, stellar masses $M_{*} \sim 10^{8.75} M_{\sun}$ (similar to the Small Magellanic Cloud). This is the first time that  weak lensing masses have been obtained for such faint dwarfs. For these faint blue dwarf galaxies, the observed SHMR deviates from simple power law extrapolations from higher masses, as well as from the predictions of abundance matching. This deviation has been noted by \citet[see Fig.\ 9]{BoyBulKap12}, who showed from dynamical measurements that low-mass galaxies ($M_* < 10^9 M_{\sun}$) lie off the predictions of abundance matching. A similar conclusion was reached by \citet{FerAbaNav12}. The latter authors note that the conflict between masses estimated via rotation curves and those predicted from abundance matching could be resolved if, for some reason, rotation velocities underestimate the circular velocities. 

\begin{figure}
\begin{center}
\includegraphics[width=\columnwidth]{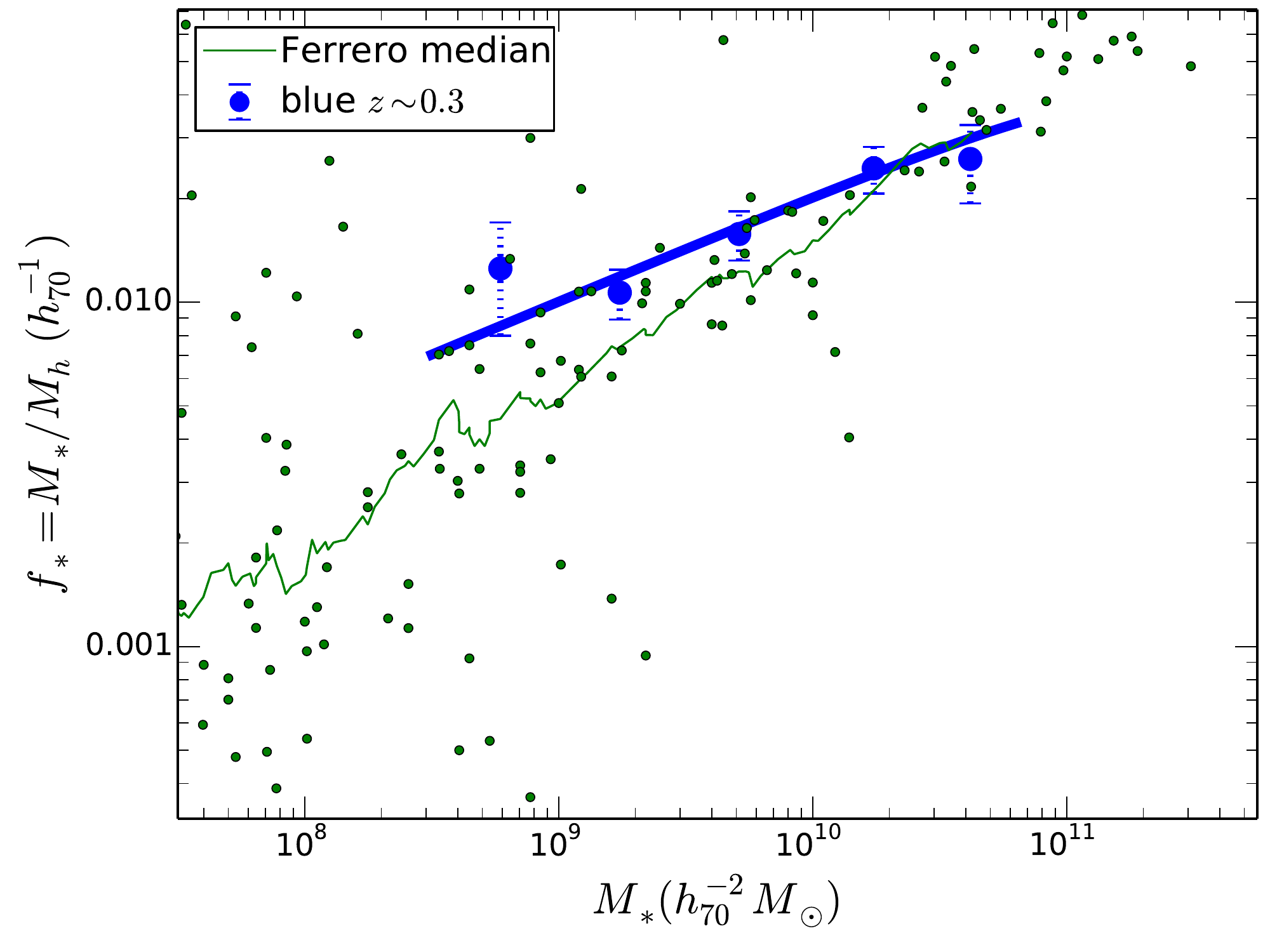}
\caption{
The SHMR for blue dwarfs at low redshift. Note that the plot extends to lower stellar masses than previous figures. The small green dots show estimates based on $z \sim 0$ rotation curves from the compilation of \protect\cite{FerAbaNav12}. The jagged green line is a running median of these data.  The blue data points with error bar and dotted blue line show the weak lensing data at $z \sim 0.3$.
}
\label{fig:faintbluedwarfs}
\end{center}
\end{figure}

In \figref{faintbluedwarfs}, we show the SHMR derived from galaxy rotation curves at $z \sim 0$, compiled by \cite{FerAbaNav12}. 
While there is considerable scatter in the SHMR from galaxy-to-galaxy, the medians of the data have a power law slope is $\beta \sim 0.5$. This is in good agreement with the mean lensing SHMR for blue galaxies, which has a power law slope $\beta = 0.44\pm0.08$ (last row of \tabref{shmrblue}). There may be a small offset of $\sim 25$\% with halo masses from lensing being slightly smaller, so it does not seem as if the problem lies on the rotation curve underestimating halo masses. 

\section{Discussion}

\label{sec:discuss}

\subsection{Understanding evolution in the SHMR diagram}
\label{sec:evolution}

\begin{figure}
\begin{center}
\includegraphics[width=\columnwidth]{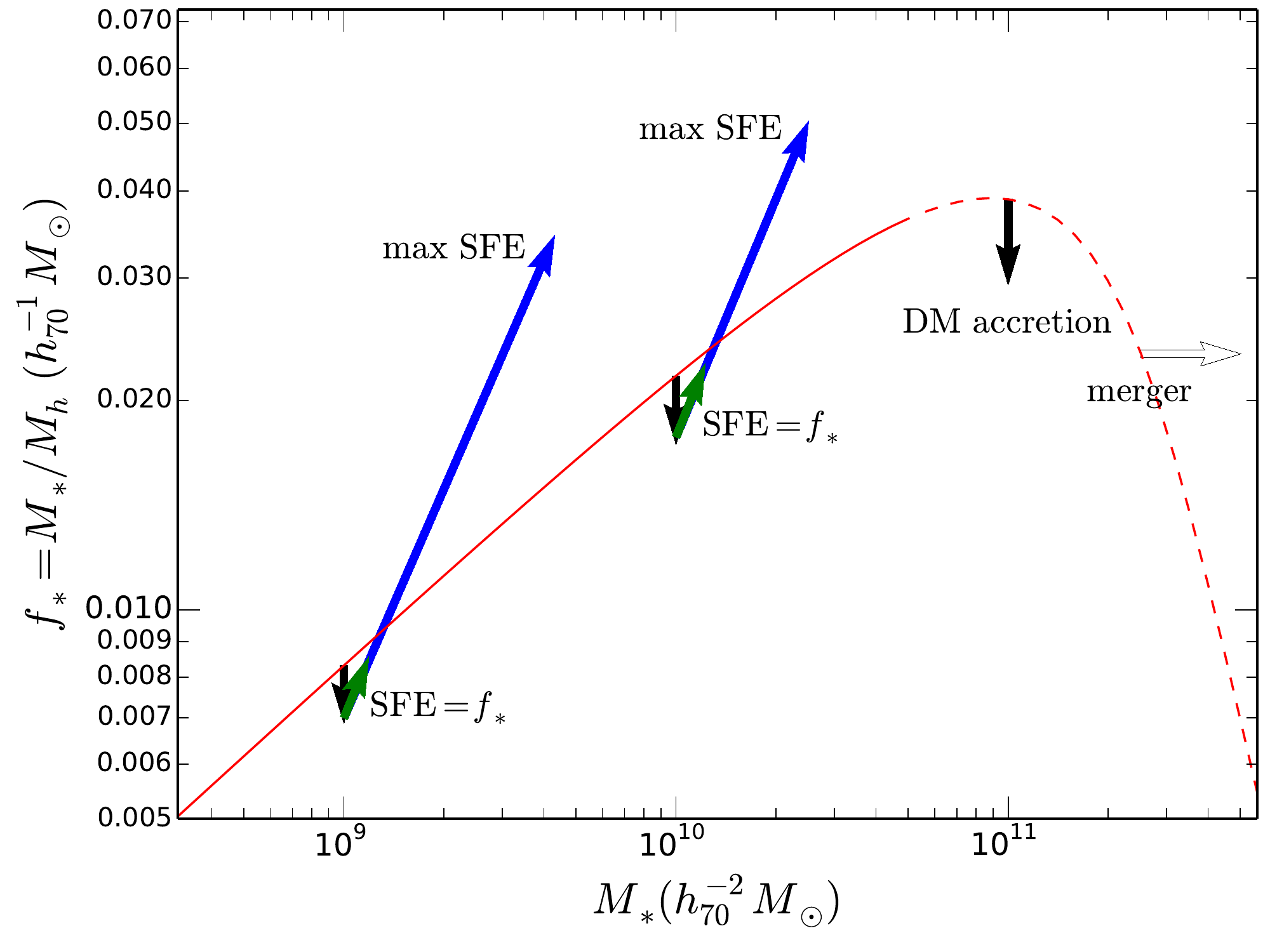}
\caption{
Physical processes that affect the evolution in the SHMR. The curve shows the SHMR at $z = 0.7$. Arrows show various processes that affect the evolution of DM or stellar mass or both, extrapolated from $z= 0.7$ to $z =0.3$. The black arrows show the effects of DM accretion based on \protect\cite{FakMaBoy10}, for haloes of three different masses. The diagonal coloured arrows show the effect of star formation: the blue arrow shows the maximal amount of star formation, i.e. one where all accreted baryons are converted to stars. 
In this case, the SHMR relation itself evolves to higher $f_*$ at given \mstel. The green arrow shows the effect of star formation assuming the efficiency is the same as $f_*$. In the latter case, the net effect of DM accretion and star formation is to move the galaxy to higher stellar mass at the same $f_*$. The white arrow shows the effect of the merger of two identical galaxies in identical haloes: $f_*$ is unchanged but the stellar mass increases.
}
\label{fig:arrows}
\end{center}
\end{figure}

The SHMR diagram is the ratio of stellar to DM mass, as a function of stellar mass. Therefore any process that affects either the stellar mass or the dark matter halo mass will move the location of a galaxy in this diagram.  The first such process is dark matter accretion: dark matter haloes, provided they are `centrals' and not `satellites' or subhaloes, will  accrete matter from their surroundings, either smoothly or from mergers of smaller haloes. This is well understood from N-body simulations in the \lcdm\ model. The downward pointing arrows in \figref{arrows} show the effect of dark matter accretion, based on the mean mass accretion history of \citet[eq.\ 2]{FakMaBoy10}, which is a function of halo mass and redshift.

Star formation creates stellar mass and moves a galaxy in \figref{arrows} on an upward diagonal line with slope one. In the current picture of galaxy formation, blue galaxies progress along a star-forming sequence, with decreasing specific star formation rates as their stellar masses increase \citep{BriChaWhi04}. The sequence itself, i.e. the specific star formation rate at a given stellar mass also evolves towards lower specific star formation rates as a function of increasing cosmic time \citep{NoeWeiFab07}.  Therefore, we expect blue galaxies that are star-forming to evolve in the SHMR diagram. The blue galaxies are almost all central galaxies, and so they will also accrete dark matter. Therefore whether these galaxies move to the right of the SHMR locus, along the locus or above it depends on the balance between the star formation rate and the dark matter halo accretion rate. Three scenarios are plotted in \figref{arrows}.

Once star formation is quenched, there are two possibilities: either a galaxy becomes a satellite, or it remains a central. In the former case, we expect the dark matter halo to be stripped, in which case the ratio $f_{*}$ should increase, provided the denominator $M_{h}$ is the actual dark matter halo mass. However, in the analysis in this paper, the predicted $\DelSig$ already assumes that the satellites have been partially stripped (equation \ref{eq:trunc}) and so the fitted parameter $M_{200}$ actually represents the pre-stripped mass.  Therefore, we expect no change due to stripping given our definition of $f_{*}$.

If the red galaxy is a central galaxy, then the dark matter halo will continue to grow by accretion of dark matter and haloes. For the evolution of the stellar mass, there are two possibilities. If the galaxies in the accreted haloes become satellite galaxies, then the stellar mass of the central galaxy remains unchanged and so the ratio $f_{*}$ will decrease as their stripped halo mass is added to the central DM halo.  On the other hand, if these galaxies merge with the central, then this will boost the stellar mass of the central. For example, if two identical galaxies in identical haloes merge, they will both be combined into a single point that is shifted horizontally to the right by $\log_{10}(2) = 0.301$ in \figref{arrows}. Of course, in reality, nearly all mergers will be less than 1:1 in mass ratio so the effect will be smaller, and in general, will not be of two galaxies with equal initial $f_{*}$.

\subsection{Towards a physical model for SHMR evolution}

The fitted SHMR and its evolution, presented in \secref{shmr}, is a purely parametric model without a physical basis. As discussed above, we can model some of the physical processes that move a galaxy in the SHMR diagram as a function of time. While a galaxy is on the blue sequence, the dominant processes are dark matter accretion and star formation.  While it is forming stars it must move to the right in the diagram, but as discussed above, how much it moves vertically depends on the balance between star formation and dark matter accretion. At some point, star formation is quenched. Observations suggest that, at least at the high masses studied here, the dominant quenching process is not environmental but rather ``internal'' to the galaxy itself \citep{PenLilKov10}.    

The star formation rates of star-forming galaxies have been well-studied empirically.  In most fits, star formation rate is a function of stellar mass and redshift. As a fiducial model, we adopt the star formation model of \cite{GilBowGla11}. 
We assume that a fraction 0.6 of the newly-formed stars are retained as long-lived stars after stellar mass loss \citep[supernovae, stellar winds;][]{BalGlaDri08}. 
The quenching mechanism may depend simply on the stellar or halo mass of the galaxy, or a different property such as the star formation rate \citep{PenLilKov10} or stellar density.  The ``downsizing'' phenomenon suggests that it may also depend explicitly on redshift.  As an example, we model quenching as a simple stellar-mass-dependent and redshift-dependent function. \cite{MouCoiAir13} find that the transition or crossover mass (where the number of red and blue galaxies is equal, or, equivalently, where the quenched fraction is 0.5) scales with redshift as $(1+z)^{1.5}$ and has a value $10^{10.75} M\sbr{\sun}$ at $z = 0.7$, consistent with \cite{PozBolZuc10}.

\begin{figure}
\begin{center}
\includegraphics[width=\columnwidth]{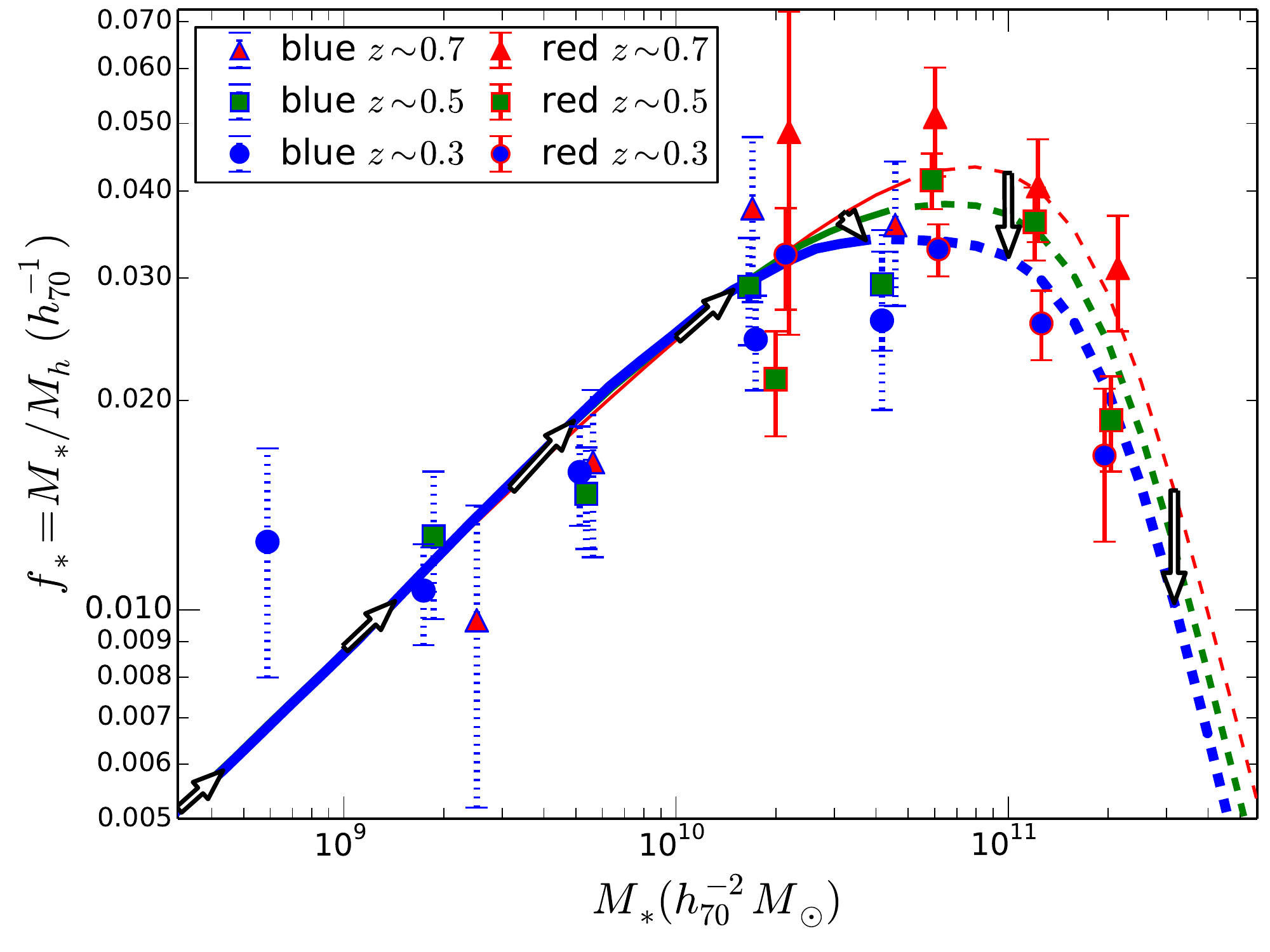}
\caption{
SHMR data compared to a model in which star formation follows the empirical star formation prescription and an empirical quenching prescription (see text for details). Large arrows show the evolutionary tracks of individual galaxies of different stellar masses as they evolve from $z=0.7$ to $z=0.3$. Notice that blue galaxies evolve \emph{along} the SHMR relation. Red galaxies have a decreasing $f_{*}$, consistent with that expected from pure dark matter accretion.}
\label{fig:physical}
\end{center}
\end{figure}

Since we have no physical model for the initial ($z = 0.7$) SHMR, this is fit with a M10 double power law.  Galaxies more massive than $M_{*} \sim 10^{10.75} M_{\sun}$ are assumed to be quenched. Subsequent evolution to $z= 0.5$ and $z=0.3$ is given by the dark matter accretion, star formation and quenching prescriptions described above.  This model therefore has only four free parameters, fewer than the parametric fits in \secref{results}. The results are shown in \figref{physical}. Overall, the model reproduces the trends seen in the data. The fit has $\chi^{2} = 20$, statistically equivalent to the best parametric models in \tabref{shmrall}, given that this model has 2 fewer free parameters.

The evolution of star forming galaxies is particularly interesting. The star formation rates from \cite{GilBowGla11} balance the mean dark matter halo accretion rates from \cite{FakMaBoy10} in such a way that galaxies evolve mostly \emph{along} the SHMR relation, with only a small amount of vertical offset that is consistent with the observational uncertainties. There is no \emph{a priori} reason that these two functions had to balance in just this way.
Thus the evolution of the SHMR can be used to understand the mean star formation history.

\section{Conclusions}
\label{sec:conc}

The depth and area of the CFHT Legacy Survey has allowed us to study the relationship between stellar and halo mass in red and blue galaxies over a wider range of stellar mass and redshift than was heretofore possible with weak lensing. The main conclusions are:
\begin{enumerate}

\item From weak lensing alone, we confirm that the SHMR peaks at halo masses $M_{h} = 10^{12.23\pm0.03} M_{\sun}$ with no significant evolution in the peak halo mass detected between  redshifts $0.2<z<0.8$.

\item The SHMR does evolve in the sense that the peak stellar-to-halo mass ratio drops as a function of time.
This result is formally statistically significant at the 99.99\% confidence level for our parametric model. 
As the peak halo mass remains constant, this means that it is the peak stellar mass which is evolving towards lower stellar masses as the galaxy redshift decreases. This is consistent with a simple model in which the stellar mass at which galaxies are quenched evolves towards a lower stellar mass with time.

\item The population of blue galaxies does not evolve strongly in the SHMR diagram.  This implies that their star formation balances their dark matter accretion so that individual galaxies move along the SHMR locus with cosmic time.

\item For the first time, weak lensing measurements of the halo mass extend to blue dwarf galaxies as faint as $M_{r} \sim -18$, with stellar masses comparable to the Magellanic Clouds. The relatively flat power law of the SHMR of blue galaxies as a function of stellar mass that was noted previously via studies of rotation curves is present in the weak lensing SHMR as well.
\end{enumerate}

\section*{Acknowledgments}

We thank  Alexie Leauthaud, Rachel Mandelbaum and Ismael Ferrero for providing their data tables in a convenient format.

MJH acknowledges support from NSERC (Canada).
H. Hildebrandt is supported by the Marie Curie IOF 252760, by a CITA National Fellowship, and the DFG grant Hi 1495/2-1.
CH acknowledges support from the European Research Council under the EC FP7 grant number 240185.
H. Hoekstra acknowledges support from  Marie Curie IRG grant 230924, the Netherlands Organisation for Scientific Research (NWO) grant number 639.042.814 and from the European Research Council under the EC FP7 grant number 279396.
TDK acknowledges support from a Royal Society University Research Fellowship.
YM acknowledges support from CNRS/INSU (Institut National des Sciences de l'Univers) and the Programme National Galaxies et Cosmologie (PNCG).
LVW acknowledges support from the Natural Sciences and Engineering Research Council of Canada (NSERC) and the Canadian Institute for Advanced Research (CIfAR, Cosmology and Gravity program).
BR acknowledges support from the European Research Council in the form of a Starting Grant with number 24067. 
TS acknowledges support from NSF through grant AST-0444059-001, SAO through grant GO0-11147A, and NWO.
LF acknowledges support from NSFC grants 11103012 \&11333001, Innovation Program 12ZZ134 of SMEC, STCSM grant 11290706600, Pujiang Program 12PJ1406700  \& Shanghai Research grant 13JC1404400.
ES acknowledges support from the Netherlands Organisation for Scientific Research (NWO) grant number 639.042.814 and support from the European Research Council under the EC FP7 grant number 279396.
CB is supported by the Spanish Science Ministry AYA2009-13936 Consolider-Ingenio CSD2007-00060, project2009SGR1398 from Generalitat de Catalunya and by the the European Commission's Marie Curie Initial Training Network CosmoComp (PITN-GA-2009-238356).
MV acknowledges support from the Netherlands Organization for Scientific Research (NWO) and from the Beecroft Institute for Particle Astrophysics and Cosmology. 
TE is supported by the Deutsche Forschungsgemeinschaft through project ER 327/3-1 and the Transregional Collaborative Research Centre TR 33 - "The Dark Universe".

This work is based on observations obtained with MegaPrime/MegaCam, a joint project of CFHT and CEA/IRFU, at the Canada-France-Hawaii Telescope (CFHT) which is operated by the National Research Council (NRC) of Canada, the Institut National des Sciences de l'Univers of the Centre National de la Recherche Scientifique (CNRS) of France, and the University of Hawaii. This research used the facilities of the Canadian Astronomy Data Centre operated by the National Research Council of Canada with the support of the Canadian Space Agency.  We thank the CFHT staff for successfully conducting the CFHTLS observations and in particular Jean-Charles Cuillandre and Eugene Magnier for the continuous improvement of the instrument calibration and the Elixir detrended data that we used. We also thank TERAPIX for the quality assessment and validation of individual exposures during the CFHTLS data acquisition period, and Emmanuel Bertin for developing some of the software used in this study. CFHTLenS data processing was made possible thanks to significant computing support from the NSERC Research Tools and Instruments grant program, and to HPC specialist Ovidiu Toader. The early stages of the CFHTLenS project was made possible thanks to the support of the European Commission's Marie Curie Research Training Network DUEL (MRTN-CT-2006-036133) which directly supported five members of the CFHTLenS team (LF, HH, BR, CB, MV) between 2007 and 2011 in addition to providing travel support and expenses for team meetings.

{\small Author Contributions: All authors contributed to the development and writing of this paper.  The authorship list reflects the lead authors of this paper (MH, BG, JC, H. Hildebrandt) followed by two alphabetical groups.  The first alphabetical group includes key contributers to the science analysis and interpretation in this paper, the founding core team and those whose long-term significant effort produced the final CFHTLenS data product.  The second group covers members of the CFHTLenS team who made a significant contribution to either the project, this paper, or both.  The CFHTLenS collaboration was co-led by CH and LVW and the CFHTLenS Galaxy-Galaxy Lensing Working Group was led by BR and CB. }  

\appendix

\section{Corrections for bias in the photometric redshifts}
\label{sec:photozbias}

\begin{figure*}
\begin{center}
\includegraphics[width=0.33\textwidth]{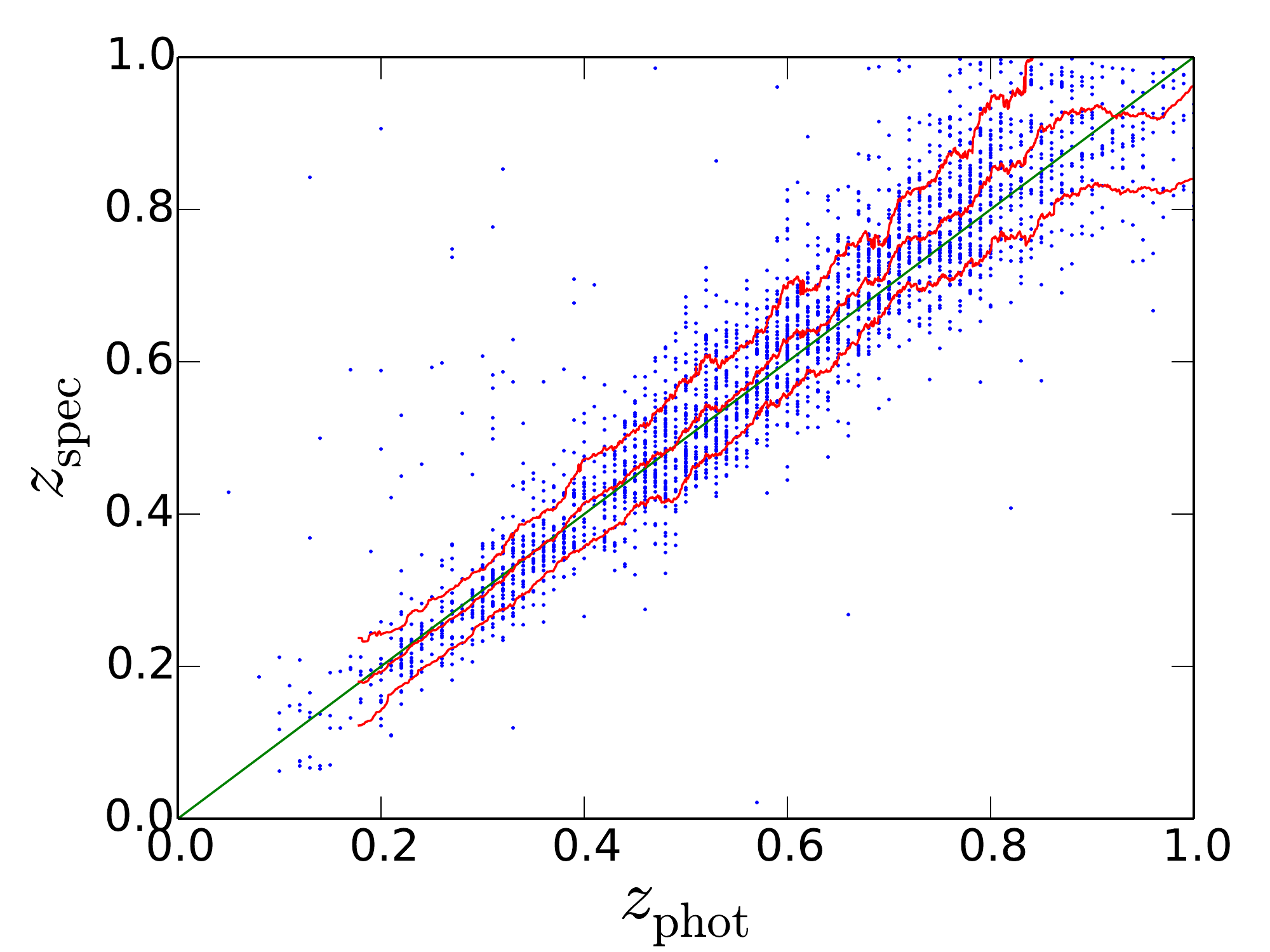}~
\includegraphics[width=0.33\textwidth]{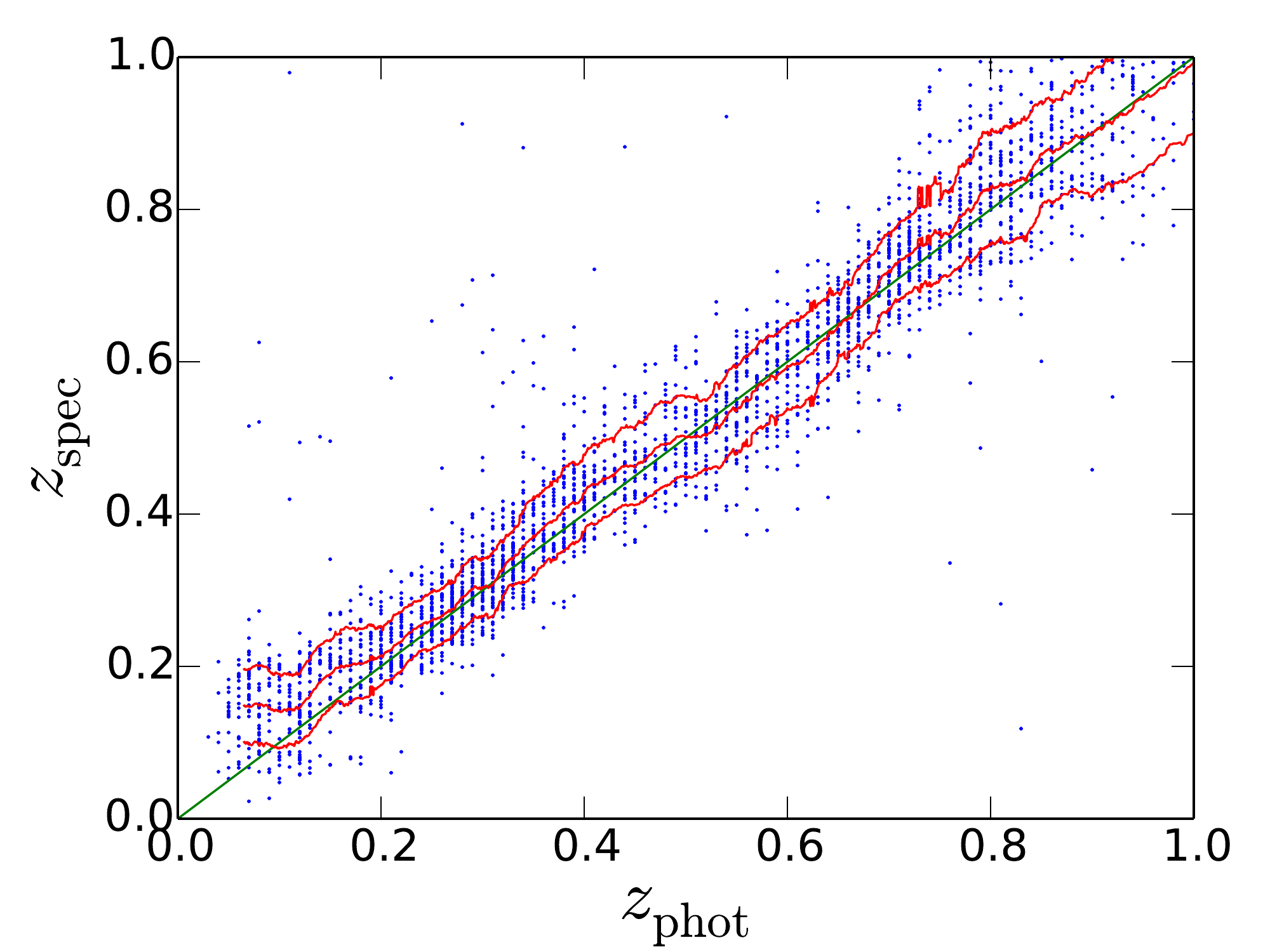}~
\includegraphics[width=0.33\textwidth]{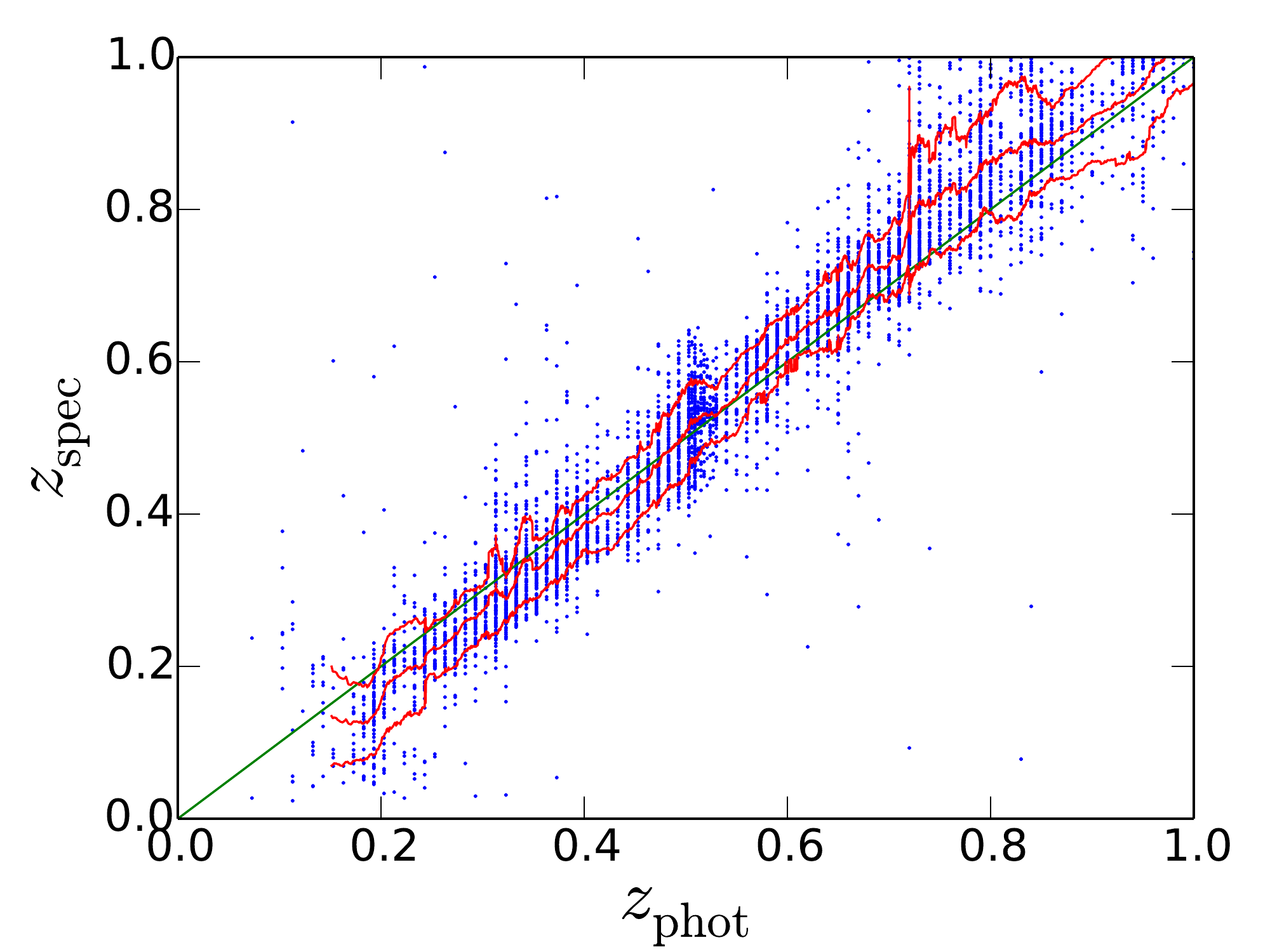}
\caption{
Spectroscopic redshifts \protect\citep[compiled by][]{HilErbKui12} as a function of photometric redshift for different spectral types. The left panel shows $T < 1.75$, the middle panel $1.75 \le T < 2.9$ and the right panel is for $T \ge 2.9$.}
\label{fig:zbias}
\end{center}
\end{figure*}

The photo-$z$'s used in this paper have small biases compared to spectroscopic redshifts (typically within the range $\pm 0.03$), as noted by \citet{HilErbKui12}. We have recomputed these biases for the range of redshifts, magnitudes and spectral types used here.  The comparison are shown in \figref{zbias}, for three spectral types $T < 1.75$,  $1.75 \le T < 2.9$ and $T \ge 2.9$.  The middle jagged red line is a running mean of $z\sbr{spec}$ as a function of $z\sbr{phot}$ after clipping $3\sigma$ outliers.  The upper and lower red lines indicate the running standard deviation (also after clipping).

The bias appears to be roughly linear function of $z\sbr{phot}$, except for the late type panel, in which there is a discontinuity at $z \sim 0.5$.  We correct this discontinuity by hand, and then fit a linear function to the clipped means to determine corrected redshifts. After correction the corrected $z\sbr{phot}$ lie within $\pm 0.01$ of $z\sbr{spec}$.  

These bias corrections are applied to the lens galaxy redshifts. We do not apply corrections to the source sample, as the latter are fainter than the spectroscopic sample.  The comparisons in \citet{HilErbKui12} show that the bias is generally not a monotonic function of magnitude, so extrapolating from the brighter lenses to the fainter sources might be dangerous.  They also show that at the faintest magnitudes, corresponding to the CFHTLenS sources, the bias appears to be small.

\section{Corrections for scatter in the photometric redshifts}
\label{sec:photozscatter}

Here we describe the corrections to the observables that are necessary because of the fact that photometric redshifts have scatter $\sim 0.04(1+z).$  As emphasised by \cite{NakManSel12} and \cite{ChoTysMor12}, this scatter leads to Eddington-like biases in several quantities of importance for weak lensing: the mean redshifts of the lenses and consequently the mean luminosities and stellar masses of the lenses, as well as the $D\sbr{ls}/D\sbr{s}$ ratio. While the CFHTLenS photo-$z$ scatter is less than the SDSS photo-$z$ scatter ($0.096(1+z)$) studied by \cite{NakManSel12}, it is nevertheless important to include these corrections.

We simulate these effects by using mock galaxy populations with known redshift and luminosity distributions, then scattering their true redshifts and re-calculating all quantities that depend on this redshift (luminosity, stellar mass, $D\sbr{ls}/D\sbr{s}$ ratio and so on). We then select lens samples in bins of (scattered) luminosity and redshift (as is done for the CFHTLenS sample) and so determine the bias in these quantitites that arises from the scatter in the photometric redshifts.  

Specifically, we select galaxies from light-cones samples of \cite{HenWhiLem12} based on the semi-analytic models of \cite{GuoWhiBoy11}. \cite{HenWhiLem12} show that these are excellent fits to the observed number counts, $n(z)$ and luminosity functions over the ranges covered by the CFHTLenS lens sample ($i'_{AB} < 23$, $0.2 < z < 0.8$).

\begin{figure}
\begin{center}
\includegraphics[width=\columnwidth]{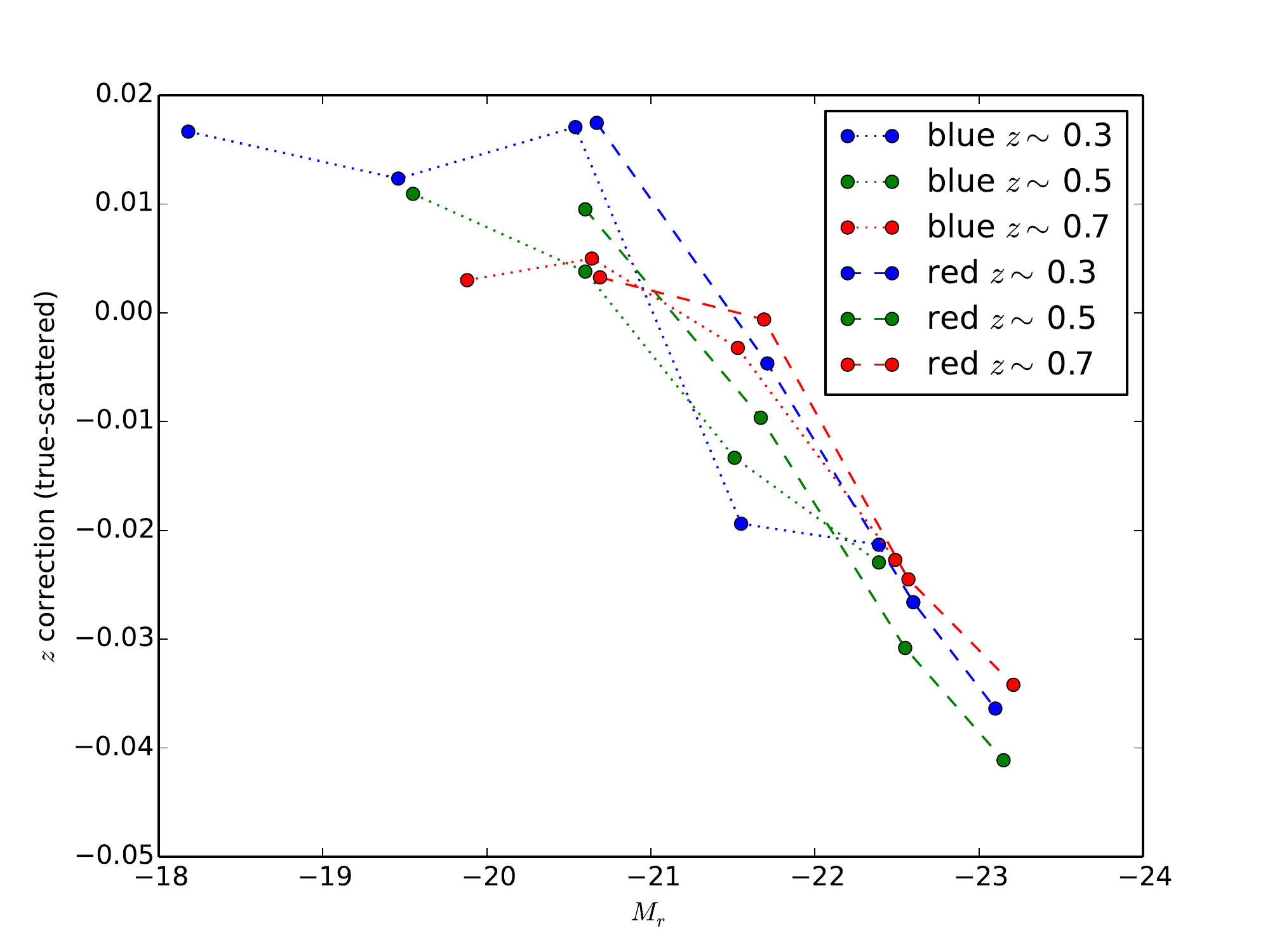}
\caption{
True lens redshift minus mean lens redshift after scattering by photo-$z$ uncertainties, as a function of (scattered) magnitude and (scattered) redshift.
}
\label{fig:zcorr}
\end{center}
\end{figure}

\begin{figure}
\begin{center}
\includegraphics[width=\columnwidth]{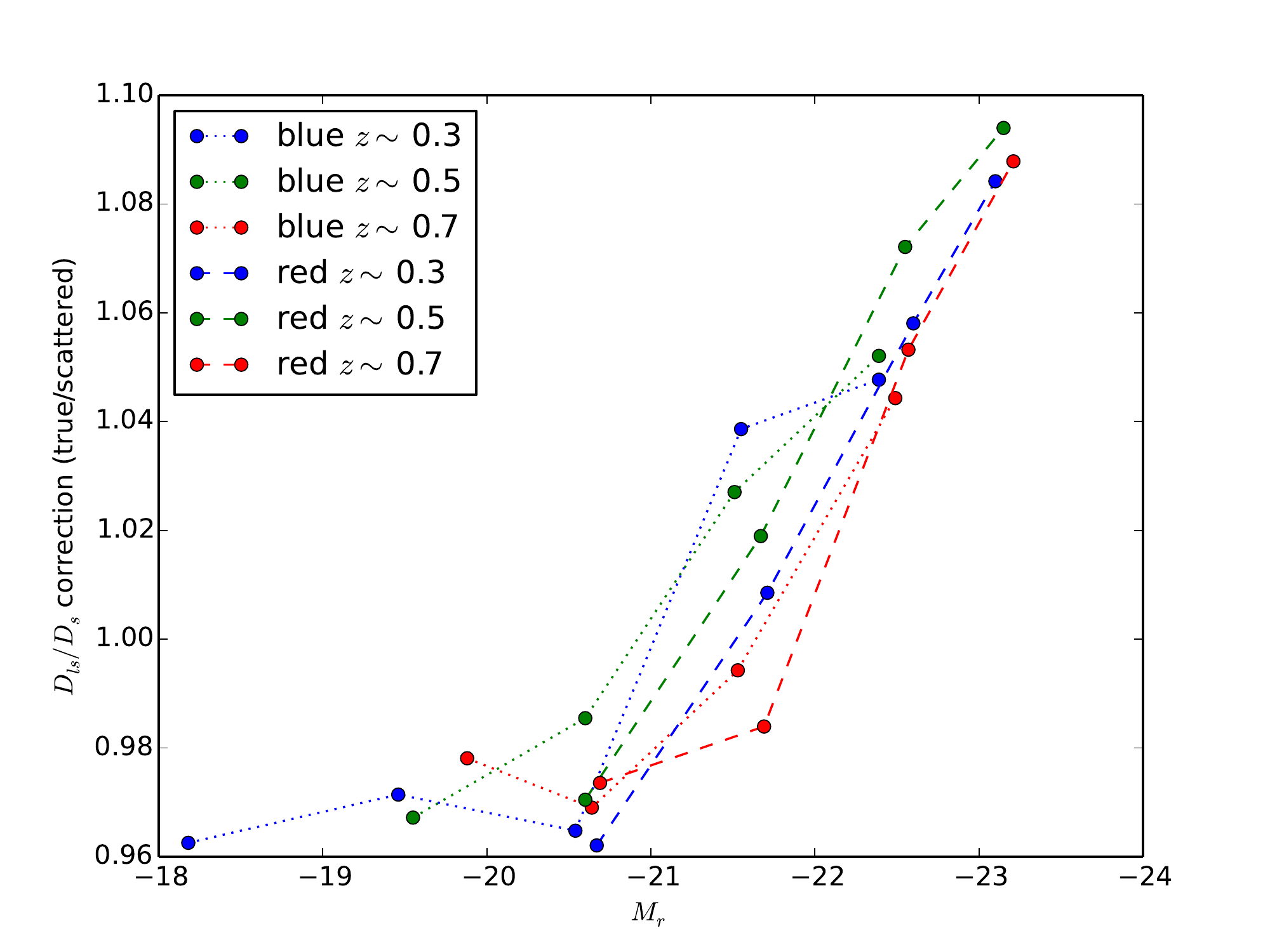}
\caption{
As in \figref{zcorr} but for the ratio of the true $D\sbr{ls}/D\sbr{s}$ ratio to its value after redshift scattering.
}
\label{fig:betacorr}
\end{center}
\end{figure}

\begin{figure}
\begin{center}
\includegraphics[width=\columnwidth]{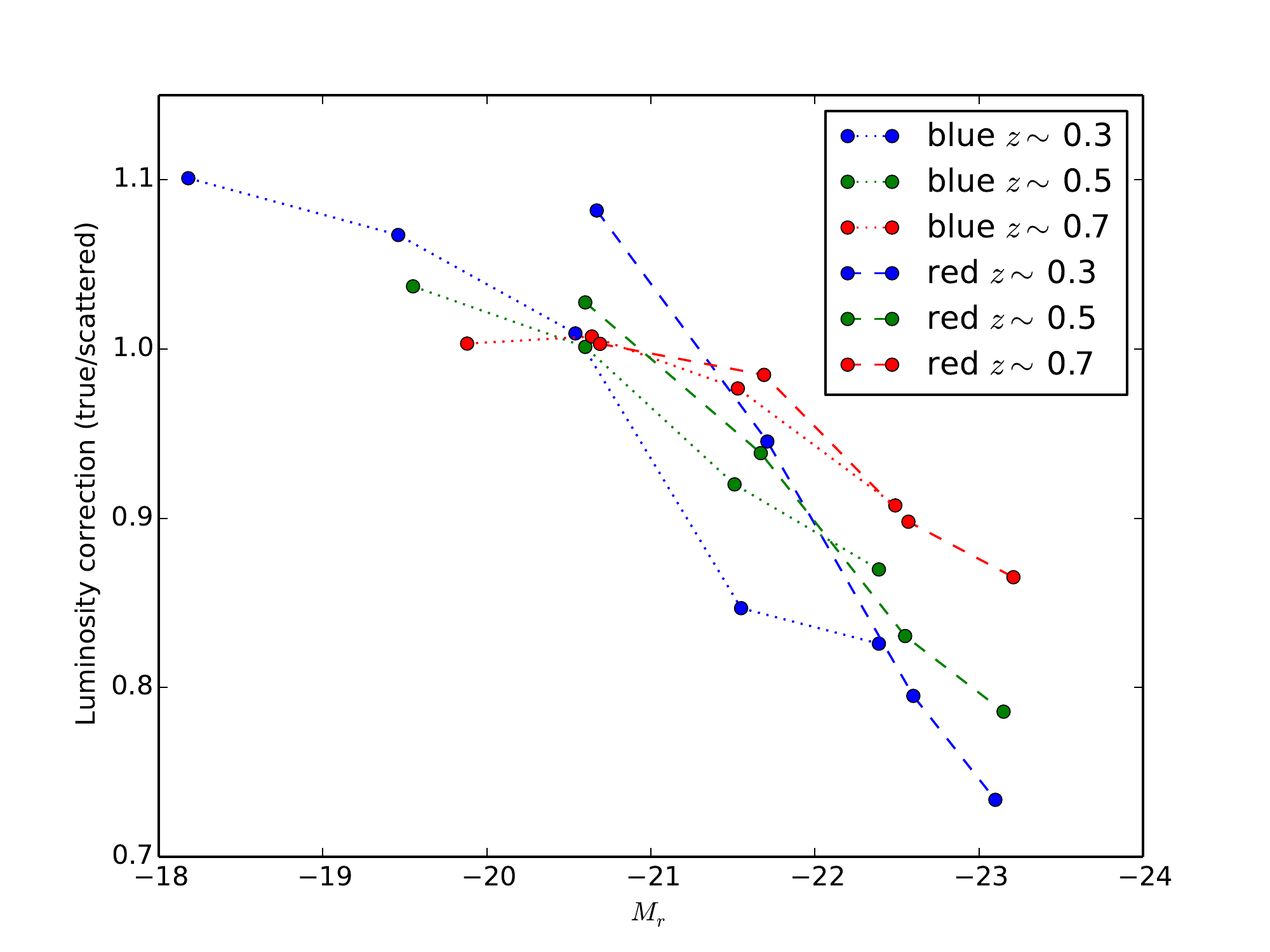}
\caption{
As in \figref{zcorr} but for the ratio of true lens luminosity to its value after redshift scattering.
}
\label{fig:lumcorr}
\end{center}
\end{figure}

The corrections derived in this way are shown in Figs. \ref{fig:zcorr}, \ref{fig:betacorr} and \ref{fig:lumcorr} for lens redshift, $D\sbr{ls}/D\sbr{s}$, and lens luminosity, respectively.

\section{SHMR as a function of $M_{*}$}
\label{sec:reverse}

For consistency and ease of comparison with previous work, the fits presented in \secref{shmr} are based on a parametrisation of $f_{*}$ as a function of halo mass, $M\sbr{h}$.  For many practical purposes \citep[e.g.][]{HudHarHar14}, however, one wants instead the halo mass as a function of stellar mass. While it is possible to use \equref{mstelscat} to obtain $M_{*}$ as a function of $M\sbr{h}$ and invert this numerically, this requires knowledge of the halo mass function and so it is more complicated to implement numerically. 

In order that readers may easily obtain the halo mass as a function of stellar mass, in this section,  we give fits to a modified version of \equref{moster}, in which the SHM ratio is a function of stellar mass $M_{*}$. Specifically, \equref{moster} is replaced with 
\begin{equation}
 f_{*}(M\sbr{*}) = 2 f_{1}^{*} \left[\left(\frac{M_{*}}{M_1^{*}}\right)^{-\beta^{*}} + \left(\frac{M_{*}}{M_1^{*}}\right)^{\gamma^{*}}\right]^{-1}\,,
\label{eq:reverse}
\end{equation} 
where the ``*'' superscript reminds the reader that the fit is based on stellar mass, not halo mass. As before we express these parameters as a function of redshift as follows
\begin{equation}
f_{1}^{*}(z) = f_{0.5}^{*} + (z - 0.5) f_{z}^{*}
\end{equation}
and
\begin{equation}
\log_{10}(M_{1}^{*})(z) = \log_{10} M_{0.5}^{*} + (z - 0.5) M_{z}^{*}\,,
\end{equation}

In this section, we fix $\gamma^{*} = 1$, for which the asymptotic behaviour is $M\sbr{h} \rightarrow$ constant as $M_{*} \rightarrow \infty$.The fits actually prefer a steeper $\gamma^{*} \sim 1.5$, but this choice of $\gamma^{*}$ would lead to a non-monotonic behaviour in which $M\sbr{h}$ rises, reaches a maximum and then declines as $M_{*} \rightarrow \infty$.  The fit with $\gamma^{*} = 1$ is only marginally worse (at the $\sim 2\sigma$ level), and avoids this undesirable, non-physical behaviour.  

\begin{figure}
\begin{center}
\includegraphics[width=\columnwidth]{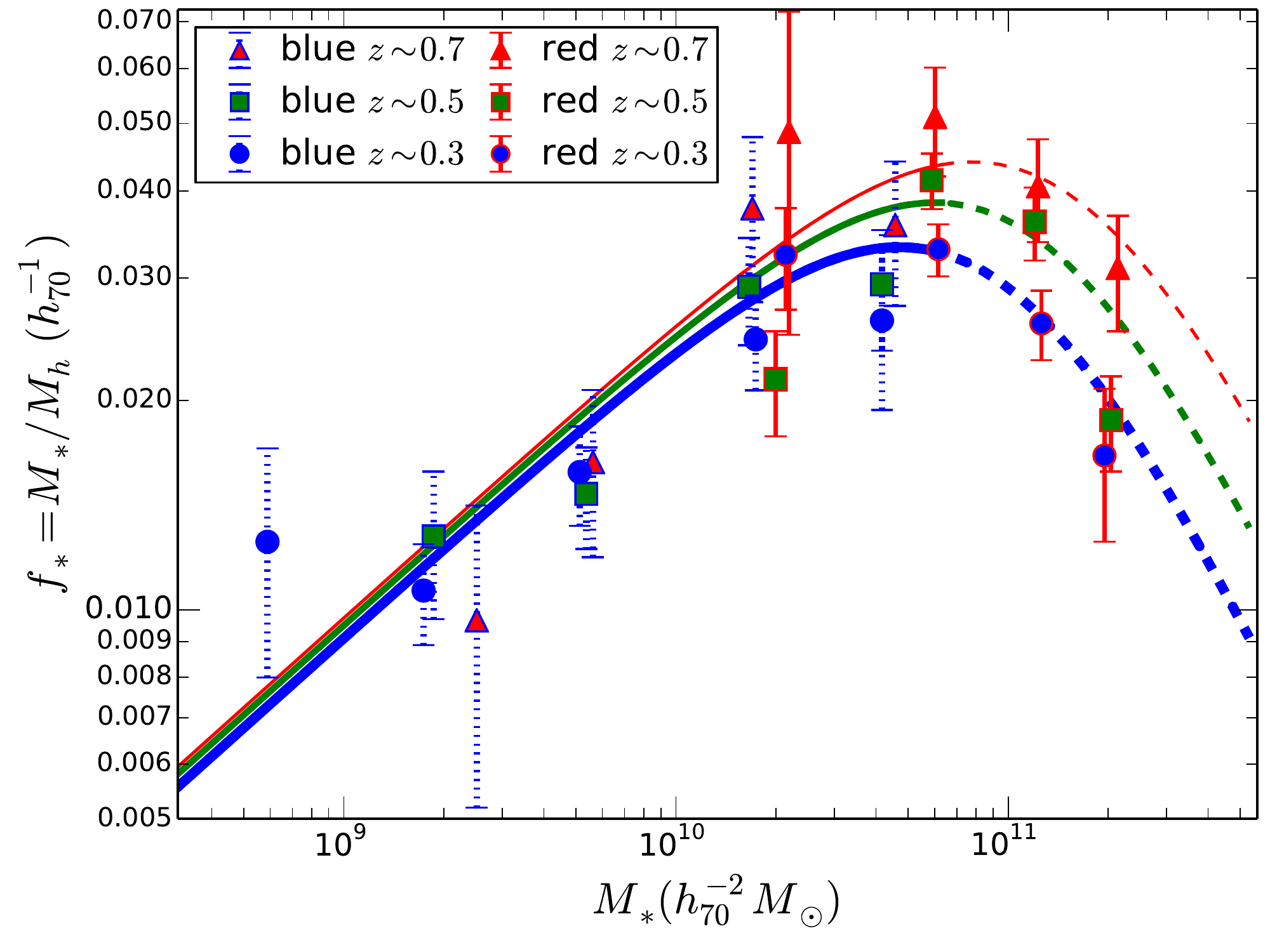}
\caption{As in \figref{shmr_all}, but the plotted lines are based on parametric form given in equation \ref{eq:reverse}, and use the ``Default'' parameters of \tabref{shmrrevall}.}
\label{fig:shmr_reverse_all}
\end{center}
\end{figure}

The results of the fit are shown in \figref{shmr_reverse_all} and \tabref{shmrrevall}.
The quality of the fit is similar to the fits as a function of $M\sbr{h}$, and qualitatively the results are similar as found in above, namely that evolution is required at a high significance level.

\makeatletter{}\begin{table*}
\newcolumntype{d}[1]{D{.}{.}{#1} }
\caption{Double power-law (M10) fits to the SHMR of all galaxies}
\begin{tabular}{lcccccccd{-1}r}
\multicolumn{1}{c}{label} & \multicolumn{1}{c}{$f_{0.5}^*$} & \multicolumn{1}{c}{$f_z^*$} & \multicolumn{1}{c}{$\log(M_{0.5}^*)$} & \multicolumn{1}{c}{$M_z^*$} & \multicolumn{1}{c}{$\beta^*$} & \multicolumn{1}{c}{$\gamma^*$} & \multicolumn{1}{c}{$\delta\sbr{blue}^*$} & \multicolumn{1}{c}{$\chi^2$} & \multicolumn{1}{c}{d.o.f.} \\
\hline
Default & $0.0357 \pm 0.0022$ & $0.026 \pm 0.009$ & $11.04 \pm 0.09$ & $0.56 \pm 0.33$ & $0.43 \pm 0.05$ & 1.0 & 0.0 & 22.0 & 20 \\
No char.\ stel.\ mass evol. & $0.0350 \pm 0.0022$ & $0.026 \pm 0.009$ & $11.05 \pm 0.09$ & 0.0 & $0.40 \pm 0.05$ & 1.0 & 0.0 & 25.0 & 21 \\
No peak $f_*$ evol. & $0.0355 \pm 0.0025$ & 0.00 & $10.98 \pm 0.09$ & $0.65 \pm 0.33$ & $0.47 \pm 0.06$ & 1.0 & 0.0 & 32.2 & 21 \\
No Evolution & $0.0340 \pm 0.0023$ & 0.00 & $11.07 \pm 0.11$ & 0.0 & $0.42 \pm 0.06$ & 1.0 & 0.0 & 36.3 & 22 \\
With Offset & $0.0369 \pm 0.0025$ & $0.026 \pm 0.008$ & $11.00 \pm 0.09$ & $0.51 \pm 0.30$ & $0.38 \pm 0.05$ & 1.0 & $-0.22 \pm 0.09$ & 17.9 & 19 \\
\label{tab:shmrrevall}
\end{tabular}
\end{table*}

\section{Treatment of stripped satellite subhalos}
\label{sec:subhalo}

In \secref{halo} we discussed our default model in which satellites have subhalos of their own. As described there, these the dark matter satellite subhalos are assumed to be partially tidally stripped.  Indeed, by comparing the weak lensing signals of satellites and field galaxies of the same stellar mass in CFHTLenS,  \cite{GilHudErb13} found that, on average, satellites had $35\pm12\%$ of their mass stripped. We assume the same stripping scheme as some previous studies (M06,V14). Thus when one stacks the lensing signal around a mixed population of a given stellar mass the predicted dark matter signal is 
\begin{multline}
\DelSig\sbr{1h, DM} = 
(1-\fsat)\DelSig\sbr{NFW}(M_{200}, c) \\
+ \fsat \DelSig\sbr{tNFW}(M_{200}, c) \,.
\end{multline}
where $\DelSig\sbr{tNFW}(M_{200}, c)$ is given by \Eqref{trunc}.

The treatment of satellite subhalos is different in some works. In particular, L12, T13 and Coupon et al. (submitted)
omit the satellite subhalo term in their lensing models, equivalent to assuming the satellites are completely stripped, or to setting $\DelSig\sbr{tNFW}(M_{200}, c) = 0$. Thus in comparison to the results presented here, for the same weak lensing measurements, the fitted mass is correspondingly increased by a factor $\sim (1+\fsat)^{-1}$. 

\begin{figure}
\begin{center}
\includegraphics[width=\columnwidth]{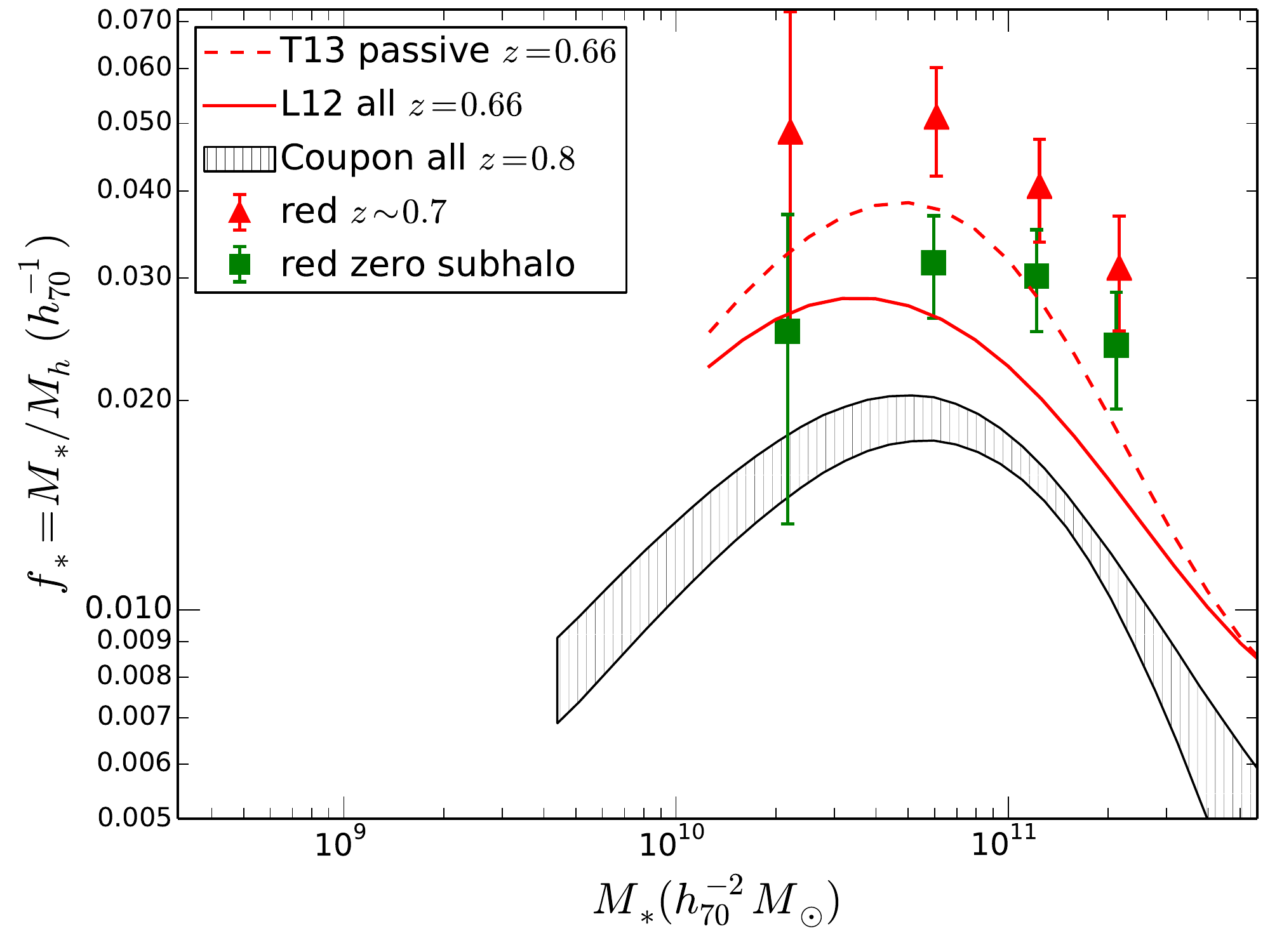}
\caption{SHMR for CFHTLens red galaxies at $z \sim 0.7$ comparing the estimated halo mass including the default assumption that satellites have partially-stripped subhalos (red triangles) and setting their subhalo contribution to zero (green squares). The latter assumption is adopted in the GGL analysis of T13, L12 and Coupon et al., whose fits are shown by the dashed red line (passive galaxies), solid red line (all galaxies) and hatched area (all galaxies), respectively.}
\label{fig:compare_sat_contrib}
\end{center}
\end{figure}

To test the effect of the treatment of satellite subhalos, we can emulate the fits of 
L12, T13 and Coupon et al. by also setting the satellite subhalo term $\DelSig\sbr{tNFW}(M_{200}, c) = 0$. The results of this modified fit for the CFHTLenS data are shown in \figref{compare_sat_contrib} for the subsample of red galaxies with $z \sim 0.7$. As expected, the modified fits yield higher fitted NFW masses and hence lower $f_{*}$ in good agreement with red galaxies from T13. The importance of this term depends on the satellite fraction, which varies from as high as 0.85 for passive low-mass satellites at high redshift to $\sim 0$ for massive active centrals.  Furthermore, the fits of L12, T13 and Coupon et al. also include abundance matching and galaxy correlation functions in addition to GGL, and indeed these may dominate over noisier GGL signal. Thus even were these authors to replicate our assumptions regarding tidal stripping, its not clear that their fits would be as affected by this choice as are our GGL-only fits. 

\footnotesize{
\bibliographystyle{mn2e}

}

\end{document}